\documentclass{article}

\usepackage{amsmath,amssymb,amsthm,bm}

\newcommand{\dbt}[1]{\ifcase#1{}\or.5\or1\or1.5\or2\or2.5\or3\or3.5\or4\or4.5\fi}

\renewcommand{\-}[1]{\mskip-\dbt#1mu}
\renewcommand\![1]{{\bm{{#1}}}}

\def\ClAuDiAeatblanks{\global\@ignoretrue}
\def\ClAuDiAstop{\relax}
\def\ClAuDiAsplit #1#2\stop{\string #1}
\def\[#1\]{\protect{\if\ClAuDiAsplit #1\stop[%
{\ClAuDiAwithlabel #1\ClAuDiAstop}\else%
{\ClAuDiAwithoutlabel #1\ClAuDiAstop}\fi}\ignorespaces}%
\def\ClAuDiAwithlabel[#1]#2\ClAuDiAstop{\protect{\begin{equation}\label{#1}\begin{split}#2\end{split}\end{equation}}}
\def\ClAuDiAwithoutlabel#1\ClAuDiAstop{\protect{\begin{equation*}\begin{split}#1\end{split}\end{equation*}}}

\newcommand{\set}[2]{\{\mskip1mu #1:#2\mskip1mu\}}

\newcommand{\sset}[1]{\bigl\{\mskip1mu#1\mskip1mu\bigr\}}

\newcommand{\defemph}[1]{{\sl #1}}

\let\temp=\colon\def\colon{\temp\mathopen{}}

\newcommand{\Ad}{\operatorname{Ad}}

\renewcommand{\Im}{\operatorname{Im}}
\renewcommand{\Re}{\operatorname{Re}}
\newcommand{\id}{\operatorname{id}}

\newcommand{\SE}[1]{\mathrm{SE}(#1)} 
\newcommand{\SO}[1]{\mathrm{SO}(#1)} 
\newcommand{\se}[1]{\mathfrak{se}(#1)}
\newcommand{\so}[1]{\mathfrak{so}(#1)}

\def\kA{{\mathcal A}}

\def\kG{{\mathcal G}}
\def\kH{{\mathcal H}}
\def\kI{{\mathcal I}}

\def\kK{{\mathcal K}}

\def\kM{{\mathcal M}}
\def\kN{{\mathcal N}}

\def\kT{{\mathcal T}}

\def\CC{{\mathbb C}}

\def\JJ{{\mathbb J}}

\def\PP{{\mathbb P}}

\def\RR{{\mathbb R}}

\def\WW{{\mathbb W}}

\def\D{{\mathrm D}}

\def\c{{\mathrm c}}
\def\d{{\mathrm d}}
\def\e{{\mathrm e}}

\def\q{{\mathrm q}}

\def\v{{\mathrm v}}

\def\fg{{\mathfrak g}}
\def\fh{{\mathfrak h}}

\def\fk{{\mathfrak k}}

\def\fq{{\mathfrak q}}

\allowdisplaybreaks

\usepackage{color}
\usepackage{picins}
\usepackage{esint}

\usepackage[pdftex]{graphicx}
\numberwithin{equation}{section}

\newtheorem{theorem}{Theorem}[section]
\newtheorem{theoremNonumber}{Theorem}

\newtheorem{proposition}[theorem]{Proposition}

\newtheorem{definition}[theorem]{Definition}
\theoremstyle{remark}
\newtheorem{remark}[theorem]{Remark}

\renewcommand{\theenumi}{\alph{enumi}}
\renewcommand{\labelenumi}{(\theenumi)}

\newcounter{badnum}
\let\oldqedsymbol=\qedsymbol
\renewcommand{\qedsymbol}{$\smash{\includegraphics[scale=.09,angle=90]{bug3.jpg}}$}
\newcommand{\LadyBirdPageBreak}{\newpage}
\newcommand{\ParagraphLadybird}[2]{\parpic{\includegraphics[scale=#2]{#1}}\noindent\ignorespaces}
\newcommand{\EndLadybird}{\bigskip\begin{center}\includegraphics[scale=.6]{bug14.jpg}\end{center}}
\newcommand{\OrnamentLadybird}[4]{\put(#1,#2){\makebox(0,0)[cc]{\includegraphics[scale=#3,angle=#4]{bug2.jpg}}}}
\newcommand{\ByeByeLadybirds}{%
  \renewcommand{\ParagraphLadybird}[2]{}%
  \renewcommand{\EndLadybird}{}%
  \let\qedsymbol=\oldqedsymbol%
  \renewcommand{\OrnamentLadybird}[4]{}
  \renewcommand{\LadyBirdPageBreak}{}%
}

\newcommand{\Agap}{\kA_e^{\rm gap}}
\newcommand{\AEM}{\kA_e^{\rm EM}}

\def\ebrk{\\&\qquad\qquad\qquad\mbox{}}

\ByeByeLadybirds
\begin{document}

\title{\vspace*{-\baselineskip}Stability Transitions for Axisymmetric 
Relative Equilibria of Euclidean Symmetric Hamiltonian~Systems\vspace*{.0in}}

\author{
George W.\ Patrick\\
\small Applied Mathematics and Mathematical Physics\\[-.05in]
\small Department of Mathematics and Statistics\\[-.05in]
\small University of Saskatchewan\\[-.05in]
\small Saskatoon, Saskatchewan, S7N~5E6, Canada\\[-.05in]
\and
Mark Roberts \& Claudia Wulff\\
\small Department of Mathematics\\[-.05in]
\small University of Surrey\\[-.05in]
\small Guildford GU2 7XH, United Kingdom\\[-.05in]
}

\date{\small January 2008}

\maketitle

\vspace*{-.3in}\begin{abstract} 
In the presence of noncompact symmetry, the stability of relative
equilibria under momentum-preserving perturbations does not generally
imply robust stability under momentum-changing perturbations. For
axisymmetric relative equilibria of Hamiltonian systems with Euclidean
symmetry, we investigate different mechanisms of stability: stability
by energy-momentum confinement, KAM, and Nekhoroshev stability, and we
explain the transitions between these. We apply our results to the
Kirchhoff model for the motion of an axisymmetric underwater vehicle,
and we numerically study dissipation induced instability of KAM stable
relative equilibria for this system.
\end{abstract}

%
\section{Introduction}
\label{sec:intro}
\ParagraphLadybird{bug13.jpg}{.15}
%
%
Relative equilibria of Hamiltonian systems with symmetry are
special solutions which are equilibria of the symmetry reduced
dynamics. Relative equilibria are called stable if they are stable equilibria
for the symmetry reduced dynamics. In other words, in a Hamiltonian
system with symmetry group $\kG$, a relative equilibrium is called
stable (more precisely $\kG$-stable [13,14,16]) if every time-orbit
close to the relative equilibrium stays close the the $\kG$-orbit of
the relative equilibrium. Since the $\kG$-orbit of the relative
equilibrium usually strictly contains its time-orbit, $\kG$-stability
is usually weaker than orbital stability. 

The stability of relative equilibria is delicate for
Hamiltonian systems with noncompact symmetry. We have
established~\cite{PRW04} exactly why it is erroneous in general to
conclude stability of relative equilibria under momentum-changing
perturbations from stability under momentum-preserving
perturbations. In the presence of noncompact symmetry, there is a gap
between these. In that gap, energy-momentum confinement fails, meaning
that  stability under momentum-changing perturbations cannot be
established by energy-momentum Lyapunov functions.

Here we investigate the stability of axisymmetric relative equilibria
of Hamiltonian systems which admit the \defemph{Euclidean symmetry}
$\kG=\SO2\ltimes\RR^3\times\SO2$. Our work bears on some results of
Leonard and Marsden~\cite{LeonardMarsden} concerning a class of
relative equilibria in the 12-dimensional Kirchhoff model for an
axially symmetric underwater vehicle, in which the vehicle falls and
spins. They derive a condition for energy-momentum confinement under
arbitrary perturbations (\cite{LeonardMarsden}, Theorem~4.4), and they
calculate that, at the boundary of this stability region, a
Hamiltonian Hopf bifurcation occurs. We show that energy-momentum
confinement under arbitrary perturbations actually occurs in a smaller
region, but that, in the intervening gap, stability can be established
by KAM methods. We also show that the coincidence of the gap boundary
and the Hamiltonian-Hopf bifurcation is to be generally expected when
the phase space is 12-dimensional.

In the underwater vehicle system, we provide numerical evidence that
the KAM stability in the gap is destroyed by small dissipation,
whereas the stability by energy-momentum confinement is preserved. In
this system, the transition from energy-momentum region to the gap is
spin independent, whereas the transition from the gap to spectral
instability, at which the Hopf bifurcation occurs, does depend on
spin. All the axisymmetric relative equilibria that are
spin-stabilized are in the gap, and their KAM stability is destroyed
by small momentum-preserving dissipation.  The implication for
gyroscopically stabilized devices is startling.  \emph{In the presence
of noncompact symmetry, robust stability may not be achievable by
the use of spin: dissipation induced loss of stability of the
relative equilibrium will occur even in absence of dissipation of
spin.}  When the symmetry is noncompact, a general understanding
of the kinds of stability which generically occur, and the transitions
between them, is necessary for the determination of robust stability
criteria.

\enlargethispage{+1\baselineskip}
This paper is structured as follows: We begin in
Section~\ref{sec:EuclAxiRE} with a description of axisymmetric
relative equilibria of Hamiltonian systems with Euclidean symmetry.
We introduce coordinates related to the reduction of the system by its
largest abelian subgroup~$\kK=\RR^3\times\SO2$ of $\kG$.  We obtain a
family of Hamiltonian systems parametrized by the corresponding
momenta. This lays plain the essential difficulty, because it shows
that \emph{perturbations to arbitrary momentum are
$\SO2$~symmetry-breaking}, for an additional $\SO2$~symmetry present
in the system which does not commute with the abelian
subgroup~$\kK$. The symmetry-breaking occurs because the additional
$\SO2$~symmetry acts on the whole family of Hamiltonian systems,
including the parameterizing momenta, and acts on one single
Hamiltonian system of the parameterized family only for certain
\defemph{vertical momenta}; only those Hamiltonian systems are
$\SO2$-symmetric.  In Section~\ref{sec:EuclStab} we study the
stability of axisymmetric relative equilibria. In general, we
establish the presence of the gap and discuss different mechanisms of
stability, and for Hamiltonian systems with 12-dimensional phase
space, we establish the Hopf eigenvalue collision at the gap
boundary. In Section~\ref{sec:appl} we begin with a brief summary of
the Kirchhoff model. We recover the stability criteria
of~\cite{LeonardMarsden} for the falling, spinning relative
equilibria, and prove stability of the relative equilibria within the
gap by verifying the Moser twist~condition for the corresponding
equilibrium on the reduced space. In Section~\ref{sec:numerics} we
numerically demonstrate that addition of small dissipation
distinguishes energy-momentum from KAM stability, by observing that
stability is maintained in the former and destroyed in the latter by
the addition of momentum conserving dissipation.

%
\section*{Acknowledgments}
\ParagraphLadybird{bug6.jpg}{.27}
%
%
This work benefited from research stays at the Banff Research Station
and the Bernoulli Center of EPFL. CW was supported by a grant from the
Nuffield Foundation and by the EPSRC First Grant Scheme. GWP is
supported by the Natural Sciences and Engineering Research Council of
Canada.

%
\section{Dynamics near axisymmetric relative equilibria}
\label{sec:EuclAxiRE}
\ParagraphLadybird{bug7.jpg}{.1}
%
%
\enlargethispage{+1\baselineskip}
In this section we introduce axisymmetric relative equilibria of
Hamiltonian systems, and introduce suitable coordinates near them for
use in the stability analysis in Section~\ref{sec:EuclStab}. We
consider a general context that includes the underwater vehicle
example in Section~\ref{sec:appl}. Let
\[[eq:Ham] 
\dot{x}=f_H(x) 
\]
be a Hamiltonian system defined by an energy $H\colon\kM\to\RR$ on
a connected symplectic manifold $(\kM,\omega)$, i.e.
\[
\omega(x)\bigl(f_H(x),w\bigr)=\D H(x)w\quad
\mbox{for all $x\in\kM,w\in\kT_x\kM$.} 
\]
Let $\!\e_1,\!\e_2,\!\e_3$ denote the unit vectors along the $x,y,z$
axis of $\RR^3$, 
and for any $v\in\RR^3$ denote
\[[eq:defhat]
v^\wedge=\widehat v=\left(\begin{array}{ccc}0&-v^3&v^2\\v^3&0&-v^1\\-v^2&v^1&0
\end{array}\right),
\]
so that, for example,
\[
\exp(\widehat{\!\e}_3\phi)=\left(\begin{array}{ccc}
\cos\phi&-\sin\phi&0\\\sin\phi&\cos\phi&0\\0&0&1\end{array}\right).
\]
Define the group
\[
\mathcal G=\SE2\times\RR\times\SO2=\SO2\ltimes\RR^3\times\SO2,
\]
where $\ltimes$ stands for a semidirect product in that the
multiplication for $g=(\phi,a,\theta)\in \SO2\ltimes\RR^3\times\SO2$
is
\[
g_1g_2=(\phi_1,a_1,\theta_1)(\phi_2,a_2,\theta_2)
=\bigl(\phi_1+\phi_2,a_1+\exp(\widehat{\!\e}_3\phi_1)a_2,\theta_1+\theta_2
\bigr).
\]
Assume $\kG$ acts symplectically on $\kM$, i.e.\ $\omega$
is $\mathcal G$-invariant, and suppose that $H$ is $\mathcal
G$-invariant. This implies that the vector field $f_H$
commutes with $\kG$, as does its flow, i.e.\ $f_H$ and its flow
are $\kG$-equivariant. Also, assume that $\kG$ acts properly and
faithfully.

In the following, $\SO2_R$ is the copy of $\SO2$ in $\mathcal G$ which
lies in the center of $\mathcal G$, so that $\theta\in\SO2_R$, and
$\SO2_L$ is such that $\phi\in\SO2_L$. Here ``$R\,$'' stands for right
and ``$L$'' for left. In the application to axisymmetric underwater
vehicles, see Section~\ref{sec:appl} below, the $\SO2_L\ltimes\RR^3$
is related to spatial isotropy, it corresponds to a left
multiplication, and it translates and rotates the body in space. The
$\SO2_R$ action is related to a material symmetry, it corresponds to a
right multiplication, and it spins the body around a symmetry axis.

By Noether's theorem there are $\dim\mathcal G=5$ locally defined
conserved quantities of~\eqref{eq:Ham}. We assume these exist globally
and organize them in the momentum map $\!J\colon\kM\to\fg^*$ such
that $\omega(x)(\xi x,w)=\D\!J_\xi(x)w$ for all $x\in\kM$, $w
\in\kT_x\kM$, $\xi\in\fg$. Here $\fg=\so2_L\ltimes\RR^3\times\so2_R=
\kT_{\id}\kG$ is the Lie algebra of $\kG$. Denote the components of
$\!J$ by
\[[eq:J]
\!J(x)=(\!J^\phi,\!J^a,\!J^\theta)(x),\quad\!J^\phi(x),\!J^\theta(x) 
\in\so2^*,\quad
\!J^a(x)\in (\RR^3)^*
\]
and the elements of $\fg^*$ by
\[
\mu=(\mu^\phi,\mu^a,\mu^\theta)\in \fg^*=\so2_L^*\ltimes\RR^3\times\so2_R^*.
\]
In the underwater vehicle example of Section~\ref{sec:appl}, the
component $\!J^a$ is the linear momentum, $\!J^\phi$ the angular
momentum, and $\!J^\theta$ the momentum of the spin of the body. As
in that system, we will assume that the momentum mapping $\!J$
transforms by the coadjoint action on $\fg^*$, i.e.~\cite{MR}
\[
\!J(g x)=(\Ad^*_{g})^{-1}\!J(x)\quad\mbox{for any $g\in\kG$ and $x\in\kM$.}
\]
Here $\Ad_g^*$ is defined by the condition
$(\Ad_g^*\mu)(\xi)=\mu(\Ad_g\xi)$, $\mu\in\fg^*$, $\xi\in\fg$,
$g\in\kG$, and $\Ad_g\xi=g\xi g^{-1}$. A standard computation (see,
for example,~\cite{MR}) gives
\[[eq:momSym]
&\!J^\phi\bigl((\phi,a,\theta)x\bigr)=\!J^{\phi}(x)
 -(\exp(\widehat{\!\e}_3\phi) a)_2\!J^{a_1}(x)
 +(\exp(\widehat{\!\e}_3\phi) a)_1\!J^{a_2}(x),\\
&\!J^{a_1}\bigl((\phi,a,\theta)x\bigr)=\!J^{a_1}(x)\cos\phi
 -\!J^{a_2}(x)\sin\phi,\\
&\!J^{a_2}\bigl((\phi,a,\theta)x\bigr)=\!J^{a_2}(x)\cos\phi
 +\!J^{a_1}(x)\sin\phi,\\
&\!J^{a_3}\bigl((\phi,a,\theta)x\bigr)=\!J^{a_3}(x),\\
&\!J^{\theta}\bigl((\phi,a,\theta)x\bigr)=\!J^{\theta}(x).
\]

%
\subsection{Axisymmetric relative equilibria of Euclidean group actions}
\label{sec:axieqintro}
%
%
\enlargethispage{+1\baselineskip}
Let $x_e$ be a relative equilibrium of~\eqref{eq:Ham}, i.e.\ let
\[
\xi_e=(\xi^\phi_e,\xi^a_e,\xi^\theta_e)\in 
\so2\oplus\RR^3\oplus\so2=\fg,
\]
and suppose $\exp(\xi_et)x_e$ is a solution of the differential
equations~$\dot x=f_H(x)$. We assume $x_e$ is \defemph{axisymmetric},
meaning that its isotropy subgroup $\kG_{x_e}=\set{g\in\kG}{gx_e=x_e}$
is the diagonal subgroup
\[
\SO2_D=\sset{(\phi,0,\phi)\in\SO2\ltimes\RR^3\times\SO2}\subseteq\kG.
\]
As is general for Hamiltonian systems with symmetry, if $\eta_e$ is in
the isotropy algebra $\fg_{x_e}$ then
$(\xi_e+\eta_e)x_e=\xi_ex_e=f_H(x_e)$ because $\eta_e x_e=0$, hence
$\exp\bigl(t (\xi_e+\eta_e)\bigr)x_e$ is also a solution of the
Hamiltonian system. Thus $\xi_e$ is only determined up to addition of
elements in
\[
\fg_{x_e}=\so2_D=\sset{(\xi^\phi,0,\xi^\phi)\in\so2\ltimes\RR^3\times\so2}
\subseteq\fg,
\] 
and so we can choose $\xi^\phi_e=0$. In the underwater vehicle, $x_e$
is an equilibrium in a frame that co-moves with the action of
$\exp(\xi_et)$,
$\omega^{\text{rot}}=\xi_e^\phi-\xi_e^\theta=-\xi_e^\theta$ is the
angular velocity, and $\xi_e^a$ is the translational velocity. The
momentum value $\mu_e=\!J(x_e)$ of the relative equilibrium is of the
form $\mu_e=(\mu^\phi_e,\mu^a_e,\mu^\theta_e)$ with
$\mu^a_e\parallel\!\e_3$, because it is fixed under the action of
$\SO2_D$ that occurs in the transformation rule~\eqref{eq:momSym}. The
momentum map $\!J^{\SO2_D}$ of the action of the symmetry group
$\SO2_D$ is
\[
\!J^{\SO2_D}(x)=\!J^\phi(x)+\!J^\theta(x).
\]

It follows from $\xi_e^\phi=0$ that $x_e$ is a relative equilibrium
for the abelian symmetry group
\[[eq:defK]
\kK=\RR^3\times\SO2_R=\sset{(0,a,\phi)\in\SO2\ltimes\RR^3\times\SO2}
\subseteq\kG,
\] 
and $x_e$ becomes an equilibrium after reduction by this group. The
stability of this equilibrium implies stability of the relative
equilibrium, see Section \ref{subsection-general-abelian-reduction}
below. We will study the stability of such an axisymmetric relative
equilibrium, which in the underwater vehicle example of
Section~\ref{sec:appl}, corresponds to a vehicle spinning about its
symmetry axis with angular velocity $\omega^{\text{rot}}$, and
translating along its symmetry axis with translational velocity
$\xi^a_e$.

%
\subsection{Reduction by $\kK=\RR^3\times\SO2_R$}
\label{subsection-general-abelian-reduction}
%
%
The following theorem provides the coordinates which we require for
the subsequent stability analysis.
\begin{theorem}\label{Th:ReduceRightSym}
In a $\kG$-invariant neighborhood of an axisymmetric relative
equilibrium $x_e$ of~\eqref{eq:Ham} there are coordinates
$x=(a,\theta,\nu^a,\nu^\theta,w)$ such that 
\begin{enumerate}
\item $\!J^a=\nu^a$, $\!J^\theta=\nu^\theta$, and the differential
equations~\eqref{eq:Ham} are
\[[eq:Bundle]
\dot a=\D_{\nu^a}H(\nu,w),\;\;\dot{\theta}=\D_{\nu^\theta}H(\nu,w),\;\;
\dot\nu^a=0,\;\;\dot\nu^\theta=0,\;\;\dot w=\JJ \D_w H(\nu,w),
\]
where $\nu=(\nu^a,\nu^\theta)$, $\JJ$ is the standard symplectic
structure matrix on the linear symplectic space $W=\RR^{\dim\kM-8}$,
and $w\in W$.
\item 
The Hamiltonian $H(\nu,w)$ is invariant under the action of $\SO2_D$,
which takes the form
\[
\nu^a\to\exp(\phi\widehat{\!\e}_3)\nu^a,
\quad\nu^\theta\to\nu^\theta,\quad w\to R_\phi(\nu,w),
\]
i.e.\ the action on $w$ generally depends on $\nu$ and this equation
defines $R_\phi$.
\item
The coordinates of the axisymmetric relative equilibrium $x_e$ are
\[
a=0,\quad \theta=0,\quad
\nu=\nu_e=\bigl(\!J^a(x_e),\!J^\theta(x_e)\bigr),\quad w=0,
\]
and $w=0$ is an equilibrium of the $\dot w$-equation
of~\eqref{eq:Bundle} at $\nu=\nu_e$.
\end{enumerate}
\end{theorem}
The $a=0$, $\theta=0$ plane is locally a slice at $x_e$ through the
action of $\kK=\RR^3\times\SO2_R$ and models the Poisson reduced
space $\kM/\kK$. The symplectic leaves of this Poisson space are given
by fixing $\nu$. At fixed $\nu$, the system
\[[eq:HamSlice]
\dot w=\JJ \D_w H(\nu,w)
\]
is the (Marsden-Weinstein) symplectic reduced system~\cite{MR}
corresponding to the abelian subgroup $\kK$ at momentum $\nu$,
obtained simply by ``ignoring cyclic coordinates''. The additional
1-dimensional symmetry is expected because $\kK$ is a codimension~1
subgroup of $\kG$. Most of the proof follows from the general theory
developed in~\cite{RWL99}, but since the symmetry group $\kK$ is
abelian there is also the following elementary proof.

\begin{proof}[Proof of Theorem \ref{Th:ReduceRightSym}]
The four momenta $\!J^a_1$, $\!J^a_2$, $\!J^a_3$, $\!J^\theta$ Poisson
commute since $\kK$ is abelian, and their derivatives are linearly
independent at $x_e$ since the isotropy group of $x_e$ is not bigger
than $\SO2_D$. Define $\nu^a=\!J^a,\nu^\theta=\!J^\theta$, and using
the Darboux theorem~\cite{W97}, choose four conjugate functions, i.e.~four
functions $a_1,a_2,a_3,\theta$, such that
\[
\{a_i,\nu^a_j\}=\delta_{ij},\quad\{\theta,\nu^\theta\}=1,\quad
\{a_i,\nu^\theta\}=0.
\]
Find functions $q_1,\ldots,q_k,p_1,\ldots,p_k$,  $k=\frac12(\dim\kM-8)$, 
that Poisson commute with the original~$8$ and satisfy
\[
\{q_i,p_j\}=\delta_{ij},\quad\{q_i,q_j\}=0,\quad \{p_i,p_j\}=0.
\]
Set $w=(q,p)$, and translate so that $x_e$ is at $w=0$. The group
$\kK$ acts by addition on the conjugate coordinates
$a_1,a_2,a_3,\theta$ since this action is generated by the momenta
$\nu^{a_1},\nu^{a_2},\nu^{a_3},\nu^\theta$. So $H$ does not depend on
the coordinates $a_1,a_2,a_3,\theta$, and we have the
$\nu=(\nu^a,\nu^\theta)$ parametrized canonical system on the linear
symplectic space $W$ given by the system~\eqref{eq:HamSlice}
of~\eqref{eq:Bundle}. The other equations of~\eqref{eq:Bundle} follow
immediately.

\enlargethispage{+1\baselineskip}
The symmetry group $\SO2_D$ acts on the variables $(w,\nu)$
independently of $a,\theta$. Indeed, if $f$ is a function which does
not depend on $a,\theta$, if $x$ is one of $a_1,a_2,a_3,\theta$ and
the conjugate of $x$ is $\bar x$, then
\begin{align}\label{eq:PoiIdent}
\frac{\partial}{\partial x}\{\!J^\theta+\!J^\phi,f\}
&=\{\bar x,\{\!J^\theta+\!J^\phi,f\}\}\notag\\
&=-\{\!J^\theta+\!J^\phi,\{f,\bar x\}\}-\{f,\{\bar x,\!J^\theta+\!J^\phi\}\}.
\end{align}
The first term of~\eqref{eq:PoiIdent} is zero because
\[
\{f,\bar x\}=-\frac{\partial f}{\partial x}=0.
\]
For the second term of~\eqref{eq:PoiIdent}, note that $\bar x$ is one
of $\nu^{a_1},\nu^{a_2},\nu^{a_3},\nu^\theta$, which are momenta of
$\kK$. Since $\kK$ is a normal subgroup of $\mathcal G$, its momenta
are a Poisson ideal of the momenta $\!J_\xi$, $\xi\in\fg$. This can
also be seen directly from~\eqref{eq:momSym} because the Poisson
brackets of $\!J^\phi$ with the other components of the momentum map
$\!J$ are found by differentiation in $\phi$ at $\phi=0$ of the right
hand sides of~\eqref{eq:momSym}. The results are linear combinations
of $\!J^\theta$ and $\!J^a_i$, $i=1,2,3$. In particular, $\{\bar
x,\!J^\theta+\!J^\phi\}$ is a linear combination of
$\nu^{a_1},\nu^{a_2},\nu^{a_3},\nu^\theta$, so the second term
of~\eqref{eq:PoiIdent} is zero for the same reason as the first. Thus
there is an action of $\SO2_D$ on the variables $(\nu,w)$. Since $\!J$
is equivariant and $\!J=(\!J^\phi,\nu)$, the resulting action on the
$\nu$ variables is independent of $w$ and equal to the coadjoint
action of $(\theta,0,\theta)$ on $\nu$, which by \eqref{eq:momSym} is
a rotation of $(\nu^a_1,\nu^a_2)$ by $\theta$. The coordinates above
can be restricted to an $\SO2_D$ invariant neighborhood because $x_e$
is fixed by the action of this compact group and are therefore
coordinates in a $\kG$-invariant neighbourhood of $\kG x_e$.
\end{proof}

\begin{definition}
A momentum $\nu\in\bigl(\RR^3\times\so2\bigr)^*$ is \defemph{vertical}
if it is fixed by the coadjoint action of $\SO2_D$, i.e.~if 
$\nu^a\parallel\!\e_3\in\mathbb R^3$. 
\end{definition}

For vertical $\nu$, the system~\eqref{eq:HamSlice} inherits an $\SO2_D$
symmetry from the full phase space $\kM$. It does not have this
additional $\SO2_D$~symmetry for nonvertical $\nu$. Hence a
perturbation of the $\SO2_D$-symmetric Hamiltonian
system~\eqref{eq:HamSlice} at a vertical momentum value $\nu$ to a
non-vertical momentum value is an $\SO2_D$-symmetry breaking
perturbation.

\begin{proposition}\label{prop:SO2Dsymmetry}
For vertical momenta $\nu$, the $\SO2_D$ action on $W$ is symplectic
with momentum map
\[
\!J_W=\!J^{\SO2_D}(a,\theta,\nu,w)\Bigr|_{a=0,\,\theta=0}.
\] 
In particular, at vertical momentum $\nu$, the
system~\eqref{eq:HamSlice} is $\SO2_D$ symmetric and conserves
$\!J_W$.
\end{proposition}
\begin{proof}
The action of $\SO2_D$ in the coordinates
$(a,\theta,\nu^a,\nu^\theta,w)$ of Theorem~\ref{Th:ReduceRightSym} is
symplectic since it is the action of a subgroup of $\kG$, which acts
symplectically by assumption. It follows that $\SO2_D$ acts
symplectically on $W$, since the symplectic form on $W$ at fixed $\nu$
is the restriction of the symplectic form on $\kM$.  Using the
symmetry properties~\eqref{eq:momSym} of the momentum map,
\[[eq:JW]
\!J_W(\nu,w)&=\!J^{\SO2_D}(0,0,\nu^a,\nu^\theta,w)\\
&=(\!J^\phi +\!J^\theta)\bigl((-a,-\theta)(a,\theta,\nu^a,
 \nu^\theta,w)\bigr)\\
&=(\!J^\phi+\!J^\theta)(a,\theta,\nu^a,\nu^\theta,w)
 +a_2\nu^a_1-a_1\nu^a_2\\
&=\!J^{\SO2_D}(a,\theta,\nu^a,\nu^\theta,w)+a_2\nu^a_1-a_1\nu^a_2,
\]
so $\!J_W=\!J^{\SO2_D}$ if $\nu_1^a=\nu_2^a=0$. Thus for fixed
vertical $\nu$, $\!J_W$ is equal to the momentum generating the
$\SO2_D$ action on~$W$.
\end{proof}

\begin{remark}
Equation~\eqref{eq:JW} and the differential equation for $\dot a$ in
the system~\eqref{eq:Bundle} implies that the conservation of $\!J_W$
is not typical for nonvertical momenta. Indeed, if $\!J_W$ is
conserved at some fixed nonvertical $\nu=\nu_0$, then adding any
function $\epsilon\tilde H(\nu,w)$ to $H$ results in
\[[eq:JWnotc]
\frac \d{\d t}(a_2\nu^a_{01}-a_1\nu^a_{02})=
\epsilon\left(\frac{\partial\tilde H}{\partial\nu^a_2}\nu^a_{01}-
\frac{\partial\tilde H}{\partial\nu^a_1}\nu^a_{02}\right).
\]
We can choose an $\SO2_D$~invariant function~$\tilde H$ such that the
right side of~\eqref{eq:JWnotc} is not everywhere zero for
$\epsilon$~arbitrarily small.
\end{remark}

\begin{remark}
Locally, the symmetry reduced space takes the form
$(\nu,w)\in\fk^*\oplus W$ where $\fk$ is the Lie algebra of $\kK$. The
space $W$ is called the {\em symplectic normal space} at $x_e$ with
respect to the symmetry group $\kK$, see~\cite{RWL99}, since it is
transversal to the group orbit ${\kK}x_e$ at $x_e$ and is a symplectic
space.  Moreover it is the largest symplectic subspace of the normal
space $\kN$ to the group orbit $\kG x_e$ at $x_e$.  In the notation
of~\cite{RWL99}, we have $\kN_0=\fk^*$, $\kN_1=W$. An application of
the results of~\cite{RWL99} would mean a reduction by the group
$\kH=\SO2_{AD}\ltimes\RR^3$ where
\[
\SO2_{AD}=\sset{(\phi,0,-\phi)\in\SO2\ltimes\RR^3\times\SO2}
\]
is the antidiagonal embedding of $\SO2$ in $\SO2_L\times\SO2_R$. But
the resulting Poisson structure on
$\kN_0=\fh^*\simeq\se2^*\oplus\RR^*$ (where $\fh$ is the Lie algebra
of $\kH$) is nontrivial since the symmetry group $\kH$ is nonabelian,
so that $\kN\simeq\kN_0\oplus\kN_1$.  In particular the symplectic
leaves have nonconstant dimension at the axisymmetric relative
equilibrium. For this reduction the conserved quantity $\!J^{\SO2_D}$
is a Poisson momentum map for the $\SO2_D$~action on
$\kN_0\oplus\kN_1$, and this momentum map is a conserved quantity for
the reduced system on $\kN_0\oplus\kN_1$. But the nontrivial Poisson
structure on $\kN_0$ would make the stability analysis more
difficult. We would have to employ the general methods developed
in~\cite{PRW04} to study stability by energy-momentum confinement and
we would have to desingularize the Poisson structure using blow up
methods as in~\cite{Patrick03} to study the KAM stability of the
axisymmetric relative equilibrium~$x_e$.
\end{remark}

\begin{remark}
In Theorem \ref{Th:ReduceRightSym}2 we could also construct an
$\SO2_D$~invariant Witt decomposition of the tangent space
$\kT_{x_e}\kM$ and then use the equivariant Darboux theorem to obtain
a model $\kK \times \fk^*\times W$ of a $\kG$-invariant neighborhood
of $\kG x_e$. In this construction the $\SO2_D$ action on $W$ would be
linear and $\nu$-independent, see~\cite{RWL99}. Note however that in
this case $\fk^*$ is not the annihilator of $\fg_{x_e}=\so2_D$, so the
situation here is not exactly the setting of~\cite{RWL99}.
\end{remark}

%
\section{Stability of axisymmetric relative equilibria}
\label{sec:EuclStab}
\ParagraphLadybird{bug11.jpg}{.5}\enlargethispage{\baselineskip}
%
%
In this section we use the coordinates from
Theorem~\ref{Th:ReduceRightSym} to study the stability of the
{\nobreak axisymmetric} relative equilibrium $x_e$. We show that
definiteness of the Hessian of $H(\nu_e,w)$ at the equilibrium~$w=0$
of~\eqref{eq:HamSlice} implies stability of~$x_e$ under arbitrary
perturbations. When this Hessian is not definite, for
momentum-preserving perturbations, a constant times the momentum
$\!J_W$ can be added to the energy to obtain a Lyapunov function of
the system~\eqref{eq:HamSlice}. However, for nonvertical linear
momentum, $\!J_W$ is not a conserved quantity of~\eqref{eq:HamSlice},
so it \emph{cannot} be used in a Lyapunov function to establish
stability for perturbations to nonvertical linear momentum.  The
\defemph{gap} (Sections~\ref{sec:intro} and~\ref{sec:KAMtheory} ) is
defined to be those relative equilibria for which a constant times the
momentum $\!J_W$ \emph{must} be added to obtain a Lyapunov
function. If $x_e$ is in the gap then the equilibrium $w=0$
of~\eqref{eq:HamSlice} has imaginary spectrum and indefinite Hessian,
and is amenable to classical KAM and Nekhoroshev results
(Section~\ref{sec:KAMtheory}) to obtain stability.

In the case of a 12-dimensional phase space $\kM$, the
system~\eqref{eq:HamSlice} has 2~degrees of freedom, and Propositions
\ref{prop:12DLinStab}--\ref{prop:HamHopf} below show that formal
stability (Section~\ref{sec:KAMtheory}) of the axisymmetric relative
equilibrium $x_e$ fails exactly at an eigenvalue collision. Moreover
we show that this is exactly where a Hamiltonian Hopf bifurcation
typically occurs and a loss of linear stability is expected. Thus, in
the case of a 12-dimensional phase space, formal stability and
linearized stability of axisymmetric relative equilibria are typically
equivalent, explaining the observation of this by Leonard and Marsden
in the Kirchhoff model of axisymmetric underwater
vehicles~\cite{LeonardMarsden}.

%
\subsection{Nonlinear Stability}
\label{sec:KAMtheory}
%
%
The relative equilibrium $x_e$ 
is called \defemph{$A$~stable} for some subset $A$ of $\kG$, if
initial data starting sufficiently close to $x_e$ stay arbitrarily
close to $Ax_e$ for all times. It is \defemph{formally stable} if
$\D^2(H+\lambda\!J_W)(x_e)|_{W}$ is definite for some
$\lambda\in\RR$. Recall from~\eqref{eq:defK} that
$\kK=\RR^3\times\SO2_R$.

\begin{proposition}\label{prop:EStabRE}
The axisymmetric relative equilibrium $x_e$ of~\eqref{eq:Ham} is
$\kK$-stable if the Hessian $\D_w^2H(\nu_e,0)$ is
definite.
\end{proposition}

\begin{proof}
If the Hessian $\D_w^2H(\nu_e,0)$ of the Hamiltonian
system~\eqref{eq:HamSlice} is definite then, using energy as a
Lyapunov function, the equilibrium $w=0$ of~\eqref{eq:HamSlice} is
stable under perturbations in $w$, i.e.~within $\nu=\nu_e$.  The
equilibrium $w=0$ persists to nearby momentum values $\nu$, and the
corresponding equilibria of~\eqref{eq:HamSlice} at those momentum
values are also stable. Since the momentum $\nu(t)$ is a conserved
quantity of~\eqref{eq:Bundle} we conclude that, for initial data close
to $\nu=\nu_e, w=0$, the solution $\bigl(\nu(t), w(t)\bigr)$
of~\eqref{eq:Bundle} stays close to $(\nu_e,0)$ for all times.  Since
the only remaining variables are $a,\theta$, and these correspond to
the action of $\kK$, the proof is complete.
\end{proof}

\begin{remark}
Better stability criteria cannot be obtained by relaxing to the weaker
stability modulo the whole group $\kG$, instead of just the subgroup
$\kK$, because the isotropy of $x_e$ implies that these two
stabilities are equivalent.
\end{remark}

Let $\!J^\kK=(\!J^a,\!J^\theta)$ denote the momentum map for the
symplectic action of $\kK$ on $\kM$.

\begin{proposition}
The axisymmetric relative equilibrium $x_e$ of~\eqref{eq:Ham} is
$\kK$-stable under perturbations that preserve the momentum
$\!J^{\kK}$ of $\kK$ if it is formally stable.
\end{proposition}

\begin{proof}
A perturbation which preserves $\!J^{\kK}$ does not change $\nu$ and
therefore is a perturbation of the stable equilibrium $w=0$ of the
system~\eqref{eq:HamSlice}.  Since $\!J_W$ is a conserved quantity of
\eqref{eq:HamSlice} at $\nu=\nu_e$ by Proposition
\ref{prop:SO2Dsymmetry}, the equilibrium $w=0$ of~\eqref{eq:HamSlice}
is stable if there is some $\lambda\in\RR$ such that $\D^2
H(0)+\lambda\D^2\!J_W(0)$ is definite.
\end{proof}

So we see that the presence of the $\SO2_D$ symmetry and its momentum
$\!J_W(w)$ at vertical momenta $\nu\parallel\!\e_3$ causes a gap
between energy-momentum confinement under momentum-preserving and
general perturbations of $x_e$.

\begin{definition}\label{df:gap}
Let $\kA_e$ be the set of axially symmetric relative equilibria. The
\defemph{EM-region} is the subset $\AEM\subseteq\kA_e$
such that $\D^2_wH(\nu_e,0)$ is definite. The \defemph{gap} is the
subset $\Agap\subseteq\kA_e$ that is formally stable but
not in the EM-region.
\end{definition}

\begin{remark}
If the symmetry group $\kG$ is
compact then the gap is absent since formal stability implies
$\kG$-stability, see e.g.~\cite{P92,PRW04}.
\end{remark}

We will say that the axisymmetric relative equilibrium $x_e$ is
\defemph{nondegenerate} if the Hessian $\D_w^2 H(\nu_e,0)$ of the
equilibrium $w=0$ of~\eqref{eq:HamSlice} is invertible. Suppose that
$x_e\in\Agap$ is nondegenerate. Then $w=0$ is a stable equilibrium
of~\eqref{eq:HamSlice}, so the linearization $\JJ\D_w^2 H(\nu_e,0)$
of~\eqref{eq:HamSlice} has nonzero purely imaginary eigenvalues,
i.e.~it is elliptic. We call the relative equilibrium $x_e$
of~\eqref{eq:Ham} \defemph{elliptic} if the corresponding equilibrium
$w=0$ of~\eqref{eq:HamSlice} is elliptic. Depending on the dimension
of the phase space~$\kM$, different scenarios are possible.
\begin{enumerate}
\item
If $\dim\kM=10$ then~\eqref{eq:HamSlice} is a 1~degree of freedom
Hamiltonian system. Equilibria of such Hamiltonian systems have
definite Hessians if and only if they are elliptic. So, the Hessian
$\D_w^2 H(\nu_e,0)$ is definite, contradicting $x_e\in\Agap$. In
particular, at this dimension there are no nondegenerate relative
equilibria in the gap.
\item
If $\dim\kM=12$ then, for each $\nu$, \eqref{eq:HamSlice} is a
2~degree of freedom Hamiltonian system. The equilibrium $w=0$ persists
as elliptic equilibrium of~\eqref{eq:HamSlice} to nearby
$\nu$. Elliptic equilibria of such Hamiltonian systems are expected to
be KAM stable~\cite{Arnold}, and so the axisymmetric relative
equilibrium $x_e$ is expected to be $\kK$-stable.
\item 
If $\dim\kM\ge14$ then the equilibrium $w=0$ persists as in case~(b),
but for each $\nu$ the system~\eqref{eq:HamSlice} has at least three
degrees of freedom. The equilibrium $w=0$ of~\eqref{eq:HamSlice} is
therefore not expected to be stable because it will exhibit Arnold
diffusion. However, Nekhoroshev stability~\cite{Fasso,Niederman} is
expected, which implies $\kK$ stability of the axisymmetric relative
equilibrium $x_e$ over exponentially long times.
\end{enumerate}

\begin{remark}
In case~(b), in order to prove KAM-stability of the axisymmetric
relative equilibrium $x_e$, one can verify the twist condition
(see~\cite{MeyerHall}) for the equilibrium $w=0$ of the 4-dimensional
$\SO2$~symmetric Hamiltonian system~\eqref{eq:HamSlice} at
$\nu=\nu_e$. If the twist condition holds at $\nu=\nu_e$, then it
holds for the equilibria persisting to nearby $\nu$ and $x_e$ is
therefore $\kK$-stable. The additional $\SO2$~symmetry
of~\eqref{eq:HamSlice} at $\nu=\nu_e$ can be used to express the
reduced Hamiltonian of~\eqref{eq:HamSlice} in terms of
$\SO2_D$-invariant functions. This simplifies the verification of the
twist condition, see Section~\ref{subsection-KAM-stability}. But, as already
emphasized, the conserved quantity $\!J_W$ of~\eqref{eq:HamSlice} at
$\nu=\nu_e$ cannot be used because it is only conserved for vertical
$\nu$. The system~\eqref{eq:HamSlice} for which the twist condition
needs to be verified, can be obtained in a concrete example by
choosing any realization of the $\RR^3\times\SO2_R$ Marsden-Weinstein
reduced space.
\end{remark}

\begin{remark}\label{rm:diss}
If $x_e\in\Agap$ then, as already noted above, $x_e$ is a spectrally
stable equilibrium with indefinite Hessian of the Hamiltonian
system~\eqref{eq:HamSlice}. As shown in~\cite{Bloch,DerksRatiu},
adding arbitrarily small dissipation to such Hamiltonian systems
results in spectral instability, because the negative eigendirection
of the Hessian forces instability in the presence of decreasing
energy. Consequently, adding a small $\SO2_D$~invariant dissipation
for each vertical~$\nu$, and extending that to nonvertical~$\nu$, will
result in dissipation induced instability at each persisting
equilibrium of the Hamiltonian systems parameterized by $\nu$. If the
dissipation preserves also the $\SO2_D$~momentum~$\!J_W$, then it
preserves all momenta resulting from the symmetries of the original
system on $\kM$. In this case, and if $D^2(H+\lambda\!J_W)(x_e)$ is
positive definite for some $\lambda$, then the persisting equilibria
of~\eqref{eq:HamSlice} remain spectrally stable for nearby vertical
$\nu$ while, for nearby nonvertical $\nu$, there is again spectral
instability. In Section~\ref{sec:numerics} we verify this by numerical
simulation of an $\RR^3\times\SO2_R$ reduction of the underwater
vehicle system.
\end{remark}

%
\subsection{Hamiltonian Hopf bifurcation}
%
%
In this section we assume that the phase space~$\kM$ is
12-dimensional, so that the system~\eqref{eq:HamSlice} is 4-dimensional.
The linearization of the action of $\SO2_D$ on $W$ at the equilibrium
$w=0$ is a linear symplectic representation, about which we need a few
elementary facts~\cite{MRS88}.  Consider a symplectic representation of
$\SO2$ on a $2d$-dimensional linear symplectic space $\WW$.  These
representations are classified by tuples of integers $n_1\le
n_2\le\cdots\le n_d$. We say that the \defemph{action is of type
$(n_1,n_2,\ldots,n_d)$} if there is a linear splitting of
$\WW$ into the sum of two $2$-dimensional invariant subspaces on which
the action is isomorphic to $z\mapsto e^{in_j\theta}z$ on the vector
space $\RR^2\cong\CC$ with its standard symplectic structure.  With
respect to such a splitting, the associated momentum mapping is
\[
\frac12 n_1|z_1|^2+\cdots+\frac12 n_d|z_d|^2,
\]
and particularly, the quadratic momentum map for type
$(n_1,n_2,\ldots,n_d)$ actions has both positive and negative definite
eigendirections if both positive and negative $n_j$ occur in its type.

Suppose $\WW$ is a linear symplectic space and $A\colon\WW\to\WW$ is
infinitesimally symplectic. Let $\pm i\omega$ be a simple complex
conjugate pair of purely imaginary eigenvalues of~$A$. The corresponding (2-dimensional) real eigenspace is invariant for the symplectic
flow $\exp(At)$, which defines a symplectic representation of $\SO2$
of either type $+1$ or type $-1$. The \defemph{Krein sign} of $\pm
i\omega$ is the sign of its type. We adopt the convention that the
frequency $\omega$ has the same sign as the Krein sign of the
eigenvalue $\pm i\omega$, so that the corresponding quadratic
Hamiltonian on the real eigenspace, with respect to a basis
with the standard symplectic structure, is
$+\frac12\omega(q^2+p^2)$. Since the frequencies are signed,
resonances between two frequencies are also signed e.g.\ there can be
$-1:1$ and $1:1$ resonances, and these are distinct cases.

\begin{remark}
Actions of $\SO2$ of type $(-1,1)$ occur frequently in dimension~4,
and this case occurs in the underwater vehicle example,
see~\eqref{eq:linearmom}. This case can be realized as $q\mapsto
R_\phi q$, $p\mapsto R_\phi p$ where $R_\phi$ is the action of $\SO2$
by counterclockwise rotations (see the proof of
Proposition~\ref{prop:HamHopf}). This is the standard $\SO2$ action on
$\RR^4$ obtained by lifting the standard action of $\SO2$ on
$\RR^2=\sset{q}$ to the cotangent bundle $\kT^*\RR^2=\sset{(q,p)}$.
\end{remark}

\begin{proposition}\label{prop:12DLinStab}
Let $x_e$ be an axisymmetric relative equilibrium of the
$\SO2\ltimes\RR^3\times\SO2$~symmetric Hamiltonian system
\eqref{eq:Ham} on a 12-dimensional phase space $\kM$. Assume that
$x_e$ is semisimple and elliptic, and let the action of $\SO2$ be of
type $n_j$ on the real $\pm i\omega_j$ eigenspace, where
$\omega_1,\omega_2$ are the two normal frequencies of the equilibrium
$w=0$.  Then $x_e$ is formally stable if and only if there is not an
$n_1:n_2$-resonance between $\omega_1,\omega_2$, i.e.~if and only if
$\omega_1n_2-\omega_2n_1\ne 0$.
\end{proposition}

Proposition~\ref{prop:12DLinStab} implies that, in 12~dimensions,
elliptic axisymmetric relative equilibria are typically formally
stable because resonance is atypical in the linearization of the
corresponding equilibrium in the reduction by $\kK$. The proof of this
Proposition follows by applying the following result to the linearized
$\SO2_D$ action at $x_e$, restricted to $W$.

\begin{proposition}\label{prop:EllipticStab}
Let $\WW$ be a 4-dimensional linear symplectic space and suppose
$\SO2$ acts linearly and symplectically on $\WW$ with
momentum map $\!J$. Suppose that $H\colon\WW\to\RR$ is $\SO2$
invariant and that $0\in\WW$ is an elliptic semisimple equilibrium
where the linearization $\JJ\D^2H(0)$ has eigenvalues $i\omega_j$,
$j=1,2$.  Suppose that $\omega_1<0$ and $\omega_2>0$, and let the
action of $\SO2$ be of type $n_j$ on the real $\pm i\omega_j$
eigenspace.  Then there is a $\lambda$ such that $\D_w^2H(0)+\lambda
\D_w^2\!J(0)$ is definite (i.e.\ $0\in\WW$ is formally stable) if
and only if
\[[eq:nonResonance]
\omega_1n_2-\omega_2n_1\ne 0.
\]
\end{proposition}

\begin{proof} 
Split $\WW=\RR^2\oplus\RR^2$ into the real eigenspaces of
$\JJ\D^2H(0)$ to the eigenvalues $i\omega_j$, choosing a basis giving
the standard symplectic structure matrix on each factor. Then for
$w=(w_1,w_2)\in\RR^2\oplus\RR^2$
\[[eq:resform]
\bigl\langle w,\bigl(\D^2H(0)+\lambda\D^2\!J(0)\bigr)w
 \bigr\rangle&=\frac{1}{2}(\omega_1 |w_1|^2+\omega_2 |w_2|^2)
 +\frac\lambda2(n_1|w_1|^2+n_2|w_2|^2)\\ 
&=\frac{1}{2}\bigl((\omega_1+\lambda n_1)|w_1|^2+(\omega_2+\lambda
 n_2)|w_2|^2\bigr).
\]
Since the action of $\SO2$ is nontrivial, one of $n_1,n_2$ is not
zero, and
\[
\lambda\mapsto(\omega_1,\omega_2)+\lambda(n_1,n_2)
\]
is a line. To show that \eqref{eq:resform} is definite, its is
sufficient that this line meets the first or third quadrant. But such
a line is contained in the second and fourth quadrant only if it
contains the origin, which is excluded by~\eqref{eq:nonResonance}.

Conversely, suppose that $\omega_1n_2-\omega_2n_1=0$. Then there is
a $\kappa$ such that $(n_1,n_2)=\kappa(\omega_1,\omega_2)$ and 
\eqref{eq:resform} becomes
\[
\bigl\langle w,\bigl(\D^2H(0)+\lambda\D^2\!J(0)\bigr)w
 \bigr\rangle=\frac{1}{2}(\kappa+\lambda)\bigl(\omega_1|w_1|^2
  +\omega_2|w_2|^2\bigr),
\]
and this is not definite for any $\lambda$ because
$\omega_1<0<\omega_2$.
\end{proof}

We now consider the case when condition~\eqref{eq:nonResonance} is
violated.

\begin{proposition}\label{prop:HamHopf}
Assume the setting of Proposition~\ref{prop:EllipticStab} again, but
let $H(\nu,w)$ depend on a parameter $\nu\in\RR$ and assume, as
before, that the $\SO2$ action is of type $(n_1,n_2)$. Assume that
{\renewcommand{\theenumi}{\roman{enumi}}\renewcommand{\labelenumi}{(\theenumi)}
\begin{enumerate}
\item the equilibrium is elliptic for $\nu<0$, $\nu\approx0$;
\item condition~\eqref{eq:nonResonance} holds for $\nu <0$, $\nu\approx0$,
but is violated at $\nu=0$.
\end{enumerate}}
Then:
\begin{enumerate}
\item 
if $n_1\neq-n_2$ then the equilibrium is linearly stable for
$\nu>0$, $\nu\approx 0$; and
\item 
if $n_1=-n_2$ and if the transversality
condition~\eqref{eq:TransversCond} holds then a Hamiltonian Hopf
bifurcation occurs and the equilibrium $0$ becomes spectrally unstable
for $\nu>0$.
\end{enumerate}
\end{proposition}

\begin{proof}
For part~(a), if $n_1\ne\pm n_2$ then the condition
$\omega_1n_2-\omega_2n_1=0$ at $w=0$, $\nu=0$ implies that the
eigenvalues $i\omega_j$ do not collide at $\nu=0$.  Therefore the
symplectic eigenvalue theorem~\cite{AM78} implies that the eigenvalues
are imaginary in a neighborhood of $\nu=0$.  If $n_1=n_2$ then
$0=\omega_1n_2-\omega_2n_1=n_1(\omega_1-\omega_2)$. Since the $\SO2$
action is assumed to be nontrivial, we have $n_1\neq 0$, and so
$\omega_1=\omega_2$. This contradicts the assumption
$\omega_1<0<\omega_2$.

%
%
%
%
%
%
%
%
%

For part~(b), so assuming $n\equiv n_2=-n_1$, there is a equivariant
Lagrangian splitting $\WW=\RR^2\oplus\RR^2$ such that the $\SO2$
action is $z\mapsto e^{in\theta}z$ on each factor $\RR^2\simeq\CC$.
The standard symplectic structure matrix blocks with respect to this
splitting, is the identity on the $(1,2)$-block, and minus the
identity on the $(2,1)$-block. Indeed, with respect to the symplectic
splitting of Proposition~\ref{prop:EllipticStab}, the basis
\begin{alignat*}{3}
&e_1=\frac1{\sqrt 2}\left(\begin{array}{cccc}1&0&0&1\end{array}\right),
\qquad &&e_2=\frac1{\sqrt 2}\left(\begin{array}{cccc}0&1&1&0
  \end{array}\right),\\
&e_3=\frac1{\sqrt 2}\left(\begin{array}{cccc}0&1&-1&0\end{array}\right),
\qquad
&&e_4=\frac1{\sqrt 2}\left(\begin{array}{cccc}-1&0&0&1\end{array}\right),
\end{alignat*}
accomplishes this, as is easily verified.  With respect to this basis,
write the linearization at $w=0$ as
\[
L=\JJ\D^2 H(0)=\left(\begin{array}{cc}a&b\\c&d\end{array}\right)
\] 
where $a,b,c,d$ are $\SO2$-equivariant $2\times 2$~matrices, which can
therefore be identified with complex numbers. Since $L$ is
infinitesimally symplectic, $d=-a^t$ and $b,c$ are symmetric,
i.e.~$d=-\bar a$ and $b,c\in\RR$, and the eigenvalues of $L$ are
\[
\lambda=i\Im(a)\pm\sqrt{-\Im(a)^2-(ad-bc)}=i\Im(a)\pm\sqrt{\Re(a)^2+bc}.
\]
So $L$ is spectrally stable if the discriminant $D=\Re(a)^2+bc$ is
negative, unstable if $D$ is positive, and an eigenvalue collision
occurs if $D=0$. Since $D=0$ when $\nu=0$ and $D<0$ for $\nu<0$,
$\nu\approx0$, $D$ changes sign under the transversality condition
\[[eq:TransversCond]
\left.\frac{\d D}{\d\nu}\right|_{\nu=0}\neq0,
\]
whereupon the equilibrium becomes spectrally unstable for $\nu>0$,
$\nu\approx0$, and undergoes a Hamiltonian Hopf bifurcation at $\nu=0$
see, e.g.~\cite{HamHopf}.
\end{proof}

%
\section{Application to the Kirchhoff model}
\label{sec:appl}
\ParagraphLadybird{bug5.jpg}{.15}
%
%
In this section we use the theory from Section~\ref{sec:EuclStab} to
analyze the stability of vertically falling, spinning relative
equilibria of a neutrally buoyant submerged axisymmetric body. In the
Kirchhoff approximation (see e.g.~\cite{Leonard97,LeonardMarsden}),
this is a Lagrangian system with configuration space $\SE3$, such
that the position and orientation of the body in configuration
$(A,b)\in\SE3$ are obtained relative to a reference body, by the
Euclidean transformation $x\mapsto Ax+b$. The reference body is chosen
with a vertical axis of symmetry, so that in the configuration
$(A,b)$, the submerged body has axis of symmetry in the direction
$A\!\e_3$. The Lagrangian is
\[[eq:LKirchhoff]
L(A,b,\Omega,v)=\frac12\Omega^tI\Omega-ml\!\e_3\cdot\Omega\times v
+\frac12v^tMv+mgl(\!\e_3\cdot A\!\e_3).
\]
Here $(\Omega,v)\in\kT\SE3$ are the body-referenced angular and
translational velocities, obtained by left translation, i.e.
$(\widehat\Omega,v)=(A,b)^{-1}(\dot A,\dot b)$, or
\[[eq:Omvdef]
\widehat\Omega=A^{-1}\dot A,\qquad v=A^{-1}\dot b.
\]
The \defemph{added inertia matrix} $I$ and \defemph{added mass matrix}
$M$ are the $3\times3$ diagonal matrices
\[
I=\left(\begin{array}{ccc}I_1&0&0\\0&I_1&0\\0&0&I_3\end{array}\right),\qquad
M=\left(\begin{array}{ccc}M_1&0&0\\0&M_1&0\\0&0&M_3\end{array}\right),
\]
and are determined from the shape and mass distribution of the
body~\cite{Leonard97}. The reference body is such that the center of
buoyancy is at the origin and the center of mass is at distance $l$
below the center of buoyancy, so the vehicle is bottom heavy when
$l>0$ and top heavy when $l<0$. We will assume, as in the physical
situation, that $I$, $M$, $m$ and $l$ satisfy
\[[eq:param]
I_1>0,\quad I_3>0,\quad M_1>0,\quad M_3>0,\quad I_1M_1-m^2l^2>0.
\]
The Lagrangian is positive definite quadratic in $\Omega,v$ under
assumptions~\eqref{eq:param}.

The system admits the symmetries of 
\begin{enumerate}
\item \defemph{spatial isotropy:} the left action of the subgroup
$\SE2\times\RR$, corresponding to translations in any direction
and rotations about the vertical; and
\item \defemph{material symmetry:} the action
of the subgroup $\SO2$ by multiplication of the inverse on the
right, corresponding to rotating the body about its axis. 
\end{enumerate}
Thus the system has the symmetry $\mathcal G$ considered in
Section~\ref{sec:EuclStab}. The point $x_e$ defined by
\[
x_e:\quad A=\id,\quad b=0,\quad\Omega=\frac{S_e}{I_3}\!\e_3,
\quad v=\frac{P_e}{M_3}\!\e_3,
\]
is an axisymmetric relative equilibrium, corresponding to the motion
where the vehicle spins at angular velocity $S_e/I_3$ and angular
momentum $S_e$ about its (vertical) symmetry axis, and translates
along that axis with vertical velocity $P_e/M_3$ and vertical momentum
$P_e$. We will apply the general stability theory developed in
Section~\ref{sec:EuclStab} to this family of relative equilibria.

%
\subsection{Hamiltonian formulation of the Kirchhoff model}
\label{sec:hamkir}
%
%
We briefly review the derivation of the Hamiltonian and the equation
of motion for the Kirchhoff model. See~\cite{Leonard97} for more
details. The conjugate variables $\Pi,P$ to $A,b$ are
\[
\Pi=\left.\frac{\partial L}{\partial\Omega}\right|_{(A,b)=(\id,0)}
 =I\Omega+ml\!\e_3\times v,\quad
P=\left.\frac{\partial L}{\partial v}\right|_{(A,b)=(\id,0)} 
 =Mv-ml\!\e_3\times\Omega,
\]
and these equations can be inverted to give
\begin{alignat*}{3}
&\Omega_1=\frac{M_1\Pi_1+mlP_2}{I_1M_1-m^2l^2},\quad
&&\quad\Omega_2=\frac{M_1\Pi_2-mlP_1}{I_1M_1-m^2l^2},\quad
&&\quad\Omega_3=\frac{\Pi_3}{I_3},\\
&v_1=\frac{I_1P_1-ml\Pi_2}{I_1M_1-m^2l^2},\quad
&&\quad v_2=\frac{I_1P_2+ml\Pi_1}{I_1M_1-m^2l^2},\quad
&&\quad v_3=\frac{P_3}{M_3}.
\end{alignat*}
The Hamiltonian is
\begin{align}\label{eq:HamUWV}
H(A,b,\Pi,P)&=\langle\Omega,\Pi\rangle+\langle v,P\rangle-L\notag\\
&=\frac12\langle\Omega,\Pi\rangle+\frac12\langle v,P\rangle
 -mgl(\!\e_3\cdot A\!\e_3)\notag\\
\begin{split}
&=
 \frac{M_1}{2(I_1M_1-m^2l^2)}\bigl(\Pi_1^2+\Pi_2^2\bigr)
 +\frac{1}{2I_3}\Pi_3^2+\frac{I_1}{2(I_1M_1-m^2l^2)}
 \bigl(P_1^2+P_2^2\bigr)\\
&\quad\mbox{}+\frac1{2M_3}P_3^2
 +\frac{ml}{I_1M_1-m^2l^2}(\Pi_1P_2-\Pi_2P_1)
-mgl\Gamma_3
\end{split}
\end{align}
where $\Gamma=A^T\!\e_3$. The equations of motion are
\[[eq:Kirchhoff]
\frac{\d \Pi}{\d t}=\Pi\times\Omega+P\times v-mgl\,\Gamma\times\!\e_3,\quad
\frac{\d P}{\d t}=P\times\Omega,\quad
\frac{\d \Gamma}{\d t}=\Gamma\times\Omega.
\]
This system is Poisson with bracket
\[[eq:JKirchhoff]
\{f,k\}=\nabla f^t\,\JJ(\Pi,P,\Gamma)\,\nabla k
\quad\mbox{where}\quad
\JJ(\Pi,P,\Gamma)=\left(\begin{array}{ccc}
 \Pi^\wedge&P^\wedge&\Gamma^\wedge\\
 P^\wedge&0&0\\
 \Gamma^\wedge&0&0\end{array}\right).
\]
The momentum map $\!J=(\!J^\phi,\!J^a,\!J^\theta)$, c.f.~\eqref{eq:J},
is given by
\[[eq:MomKirchhoff]
&\!J^\phi(A,b,\Pi,P)=(A\Pi+b\times AP)\cdot\!\e_3,\\
&\!J^a(A,b,\Pi,P)=AP,\\
&\!J^\theta(A,b,\Pi,P)=-\Pi\cdot\!\e_3.
\]

%
\subsection{Reduction by $\RR^3\times\SO2_R$}
\label{Sect:RightRed}
%
%
As shown in Section~\ref{sec:EuclStab}, the stability of the relative
equilibrium $x_e$ can be established by a study of the stability of
the equilibrium of~\eqref{eq:HamSlice} at the parameter value
$\nu=(\mu^a_e,\mu^\theta_e)=(P_e\!\e_3,S_e)$. The
system~\eqref{eq:HamSlice} is the Marsden-Weinstein
reduction~(see~\cite{MR}) of the full Hamiltonian system by the group
$\RR^3\times\SO2_R$, at the momentum level $(\mu^a_e,\mu^\theta_e)$,
i.e.~\eqref{eq:HamSlice} is obtained from the full Hamiltonian
system~\eqref{eq:Ham} by fixing the values of the conserved quantities
$\!J^a$ and $\!J^\theta$ to their respective values $\mu^a_e$ and
$\mu^\phi_e$, and eliminating variables along the group
$\RR^3\times\SO2_R$. For the Kirchhoff model,
equations~\eqref{eq:Kirchhoff} cannot be directly used because they
are the result of a reduction by the group $\SO2_L\ltimes\RR^3$. In
this section we compute the reduced spaces near $x_e$, in a way that
relates to the variables $\Pi,P,\Gamma$ of~\eqref{eq:Kirchhoff}, by
comoving with the action of $\SO2_R$.

Given $q_1,q_2$,
define $q=q_1\!\e_1+q_2\!\e_2$ and
\[[eq:Aphi]
A_{q_1,q_2}=\id+\widehat x
 +f\widehat x^2\Bigr|_{x=-\!\e_3\times q}
 \quad\mbox{where}\quad
 f=\frac1{1+\sqrt{1-\|q\|^2}},
\]
where $\widehat x$ is defined in \eqref{eq:defhat}. The matrix
$A_{q_1,q_2}$ in~\eqref{eq:Aphi} is orthogonal because $f$ satisfies
the equation $\|q\|^2f^2-2f+1=0$ and $f$ is the solution of this
quadratic that is smooth at $q=0$. The choice $\!x=-\!\e_3\times q$ is
so that
$A_{q_1,q_2}^t\!\e_3=q_1\!\e_1+q_2\!\e_2+\sqrt{1-q_1^2-q_2^2}\,\!\e_3$. The
map $(q_1,q_2,\theta)\mapsto A_{q_1,q_2}\exp(-\widehat{\!\e}_3\theta)$
coordinatizes $\SO3$ near the identity and provides an angle
$\theta(A)$, which may be used to define the coordinates
$(a,\theta,q_1,q_2)$ by
\[
(A,b)=(a,\theta)\cdot(A_{q_1,q_2},mlM^{-1}q)
 =(A_{q_1,q_2}\exp(-\theta\widehat{\!\e}_3),mlM^{-1}q+a).
\] 
The choice of $b$ is such that the two sections $a=0,\theta=0$ and
$q_1=q_2=0$ are orthogonal (at their intersection point) in the
kinetic energy metric. This is an example of a general procedure
advocated in~\cite{RobertsSchmahStoica}. The group $\RR^3\times\SO2_R$ acts by addition in $a,\theta$, and
\[[eq:slice_coords]
\Gamma=\exp(\theta\widehat{\!\e}_3)(q_1\!\e_1+q_2\!\e_2)
 +\Gamma_3\!\e_3
\]
so $(q_1,q_2)$ comoves with $(\Gamma_1,\Gamma_2)$.

Let the conjugate momenta to $a,\theta,q_1,q_2$ be
$\nu^a,\nu^\theta,p_1,p_2$. The coordinates $a$ and $\theta$ are
cyclic and so $\nu^a,\nu^\theta$ are conserved. Thus the
$\RR^3\times\SO2_R$ reduction results in a Hamiltonian in terms of
canonical coordinates $q_i,p_i$, which is parametrized by
$\nu^a,\nu^\theta$. This is computed in the following proposition.

\begin{proposition}\label{prop:SO2Rred}
The Hamiltonian of the $\RR^3\times\SO2_R$ reduction of the Kirchhoff
model~\eqref{eq:HamUWV} in the canonical coordinates $q_i,p_i$ is
\[[eq:HamRed]
H(q_1,q_2,\tilde p_1,\tilde p_2)&=\frac{M_1}{2(I_1M_1-m^2l^2)}\left(
 \tilde p_1^2+\tilde p_2^2-(q_2\tilde p_1-q_1\tilde p_2)^2\right)\\
&\qquad\mbox{}+\frac12
 \left(\frac1{M_3}-\frac{I_1}{(I_1M_1-m^2l^2)}\right)(\nu^a\cdot\Gamma^-)^2\\
&\qquad\mbox{}
 +\frac{ml}{I_1M_1-m^2l^2}
 \left((q_1\tilde p_2-q_2\tilde p_1)(\nu^a\cdot\Gamma^-)
 +\nu^a_2\tilde p_1-\nu^a_1\tilde p_2\right)\\
&\qquad\mbox{}-mgl\Gamma_3,
\]
where
\begin{align}
&\tilde p_1=p_2-fq_1\nu^\theta-\frac{ml}{M_1}\nu^a_2,\qquad
\tilde p_2=-p_1-fq_2\nu^\theta+\frac{ml}{M_1}\nu^a_1,\label{eq:shift}\\
&\Gamma^-=A_{q_1,q_2}^t\!\e_3
 =q_1\!\e_1+q_2\!\e_2-\Gamma_3\!\e_3,\qquad
\Gamma_3=\sqrt{1-q_1^2-q_2^2}.\notag
\end{align}
\end{proposition}
\begin{proof}
The $\RR^3\times\SO2_R$ reduced Hamiltonian can be computed by
computing $p_i$ in terms of $\Pi$ and $q_1,q_2$ and eliminating $P$
and $\Pi_3$ using (see~\eqref{eq:MomKirchhoff}) $P=A^t\nu^a$ and
$\Pi_3=-\nu^\theta$. Since the Hamiltonian is invariant, it does not
depend on $a$ or $\theta$, so we set $a=0$ and $\theta=0$. By
definition of canonical coordinates,
\[
p_i=\left(A^t\frac{\partial A}{\partial q_i}\right)^\vee\cdot\Pi
  +\left(A^t\frac{\partial b}{\partial q_i}\right)\cdot P,
\]
where the superscript $^\vee$ recovers a vector from an antisymmetric
matrix by inverting the $^\wedge$ operation defined
by~\eqref{eq:defhat}.
Then
\begin{align*}
&A_{q_1,q_2}^t\frac{\partial A_{q_1,q_2}}{\partial q_1}=\left(-\!\e_2
 +\frac{f^2q_1}{f-1}\,\!\e_3\times q +fq_2\,\!\e_3\right)^\wedge,\\
&A_{q_1,q_2}^t\frac{\partial A_{q_1,q_2}}{\partial q_2}=\left(\!\e_1
 -\frac{f^2q_2}{f-1}\,\!\e_3\times q-fq_1\,\!\e_3\right)^\wedge,
\end{align*}
from which 
\begin{align*}
p_1&=-\frac{f^2q_1q_2}{f-1}\Pi_1+\left(-1+\frac{f^2q_1^2}{f-1}\right)\Pi_2
 -fq_2\nu^\theta+\frac{ml}{M_1}\nu^a_1,\\
p_2&=\left(1-\frac{f^2q_2^2}{f-1}\right)\Pi_1+\frac{f^2q_1q_2}{f-1}\Pi_2
 +fq_1\nu^\theta+\frac{ml}{M_1}\nu^a_2.
\end{align*}
These linear equations may be solved for $\Pi_1,\Pi_2$, with the result
\[[eq:pip]
\left(\begin{array}{c}\Pi_1\\\Pi_2\end{array}\right)=
\left(\begin{array}{cc}1-fq_2^2&fq_1q_2\\fq_1q_2&1-fq_1^2\end{array}\right)
\left(\begin{array}{c}\tilde p_1\\\tilde p_2\end{array}\right)
\]
where $\tilde p_i$ are defined by~\eqref{eq:shift}. The matrix
in~\eqref{eq:pip} is the same as the upper $2\times 2$ submatrix of
$-\widehat{\!\e}_3A_{q_1,q_2}^t\widehat{\!\e}_3$, which is otherwise
sparse, so setting $\tilde p=\tilde p_1\!e_1+\tilde p_2\!\e_2$,
\[[eq:funny]
\Pi=-\!\e_3\times\bigl(A_{q_1,q_2}^t(\!\e_3\times\tilde p)\bigr)
 -\nu^\theta\!\e_3,
\]
and
\begin{align*}
&\Pi_1^2+\Pi_2^2
=\bigl\|-\!\e_3\times\bigl(A_{q_1,q_2}^t(\!\e_3\times\tilde p)\bigr)\bigr\|
=\tilde p_1^2+\tilde p_2^2-(q_2\tilde p_1-q_1\tilde p_2)^2,\\
&\Pi_1P_2-\Pi_2P_1=\!\e_3\cdot(\Pi\times A_{q_1,q_2}^t\nu^a)
=(q_1\tilde p_2-q_2\tilde p_1)(\nu^a\cdot\Gamma)
 +\nu^a_2\tilde p_1-\nu^a_1\tilde p_2,\\
&P_3^2=(\nu^a\cdot A_{q_1,q_2}^t\!\e_3)^2.
\end{align*}
The Hamiltonian~\eqref{eq:HamUWV} is obtained from~\eqref{eq:HamRed}
after substitution these and deleting inessential constants.
\end{proof}

As already noted, the case of $\nu^a$ vertical is important for the
stability and KAM analysis, and we specialize to this now.

\begin{proposition}
At vertical momentum $\nu^a_1=\nu^a_2=0$, the reduced Hamiltonian
system~\eqref{eq:HamUWV} is
\[[eq:HamRedVert]
H(q_1,q_2,p_1,p_2)
&=\frac{F_p}2
\bigl(
 (p_1+F_l\Gamma_3q_1+\nu^\theta fq_2)^2
 -(q_1p_1+q_2p_2)^2\\
&\qquad\qquad\mbox{}
+(p_2+F_l\Gamma_3q_2-\nu^\theta fq_1)^2\bigr)\\
&\qquad\mbox{}+\frac{F_q}2\|q\|^2+mgl\left(f-\frac12\right)\|q\|^2
+\frac12F_pF_l^2\|q\|^4.
\]
where
\[[eq:Fdef]
F_p=\frac{M_1}{I_1M_1-m^2l^2},\qquad 
F_q=mgl-(\nu^a_3)^2\biggl(\frac1{M_3}-\frac1{M_1}\biggr),\qquad
F_l=\frac{ml\nu^a_3}{M_1}.
\]
\end{proposition}
\begin{proof}
Using~\eqref{eq:shift}, the
Hamiltonian~\eqref{eq:HamRed} can be written in the canonical
coordinates $q_1,q_2,p_1,p_2$ as follows:
\begin{align*}
&H(q_1,q_2,p_1,p_2)\\
&\qquad=\frac{M_1}{2(I_1M_1-m^2l^2)}\left(
 \tilde p_1^2+\tilde p_2^2 -(q_2\tilde p_1-q_1\tilde p_2)^2
 \right)\\
&\qquad\mbox{}+\frac12
 \left(\frac1{M_3}-\frac{I_1}{(I_1M_1-m^2l^2)}\right)
 (\nu^a_3\Gamma_3)^2\\
&\qquad\mbox{}
 -\frac{ml}{I_1M_1-m^2l^2}
 (q_1\tilde p_2-q_2\tilde p_1)(\nu^a_3\Gamma_3)-mgl\Gamma_3\\
&=\frac{M_1}{2(I_1M_1-m^2l^2)}\left(
 \left(\tilde p_1+\frac{ml\nu^a_3}{M_1}q_2\Gamma_3\right)^2
 +\left(\tilde p_2-\frac{ml\nu^a_3}{M_1}q_1\Gamma_3\right)^2
 -(q_2\tilde p_1-q_1\tilde p_2)^2\right)\\
&\qquad\mbox{}-\frac{(\nu^a_3)^2}2
 \left(\frac1{M_3}-\frac{I_1}{(I_1M_1-m^2l^2)}
 +\frac{m^2l^2(1-\|q\|^2)}{M_1(I_1M_1-m^2l^2)}\right)\|q\|^2 -mgl\Gamma_3\\
\begin{split}
&=\frac{M_1}{2(I_1M_1-m^2l^2)}\Biggl(
 \biggl(p_1+\frac{ml\nu^a_3}{M_1}q_1\Gamma_3+\nu^\theta q_2f\biggr)^2
 -(q_1p_1+q_2p_2)^2\\
&\qquad\mbox{}
 +\biggl(p_2+\frac{ml\nu^a_3}{M_1}q_2\Gamma_3-\nu^\theta q_1f\biggr)^2
 \Biggr)\\
&\qquad\mbox{}
 +\frac12\Biggl(2mglf-(\nu^a_3)^2
 \biggl(\frac1{M_3}-\frac1{M_1}
 -\frac{m^2l^2\|q\|^2}{M_1(I_1M_1-m^2l^2)}\biggr)\Biggr)\|q\|^2,
\end{split}\end{align*} 
where a constant has been deleted at the final equality. Substitution of
\eqref{eq:Fdef} then gives the Hamiltonian~\eqref{eq:HamRedVert}.
\end{proof}

\begin{remark}\label{rem:redso2mom}
In accord with Proposition~\ref{prop:SO2Dsymmetry}, the
Hamiltonian~\eqref{eq:HamRedVert} admits an additional $\SO2$
symmetry because $\nu^a$ is vertical, which we can take as the diagonal
action of $\SO2$ on $(q_1,q_2),(p_1,p_2)$, with the standard conserved
momentum $q_1p_2-q_2p_1$. From~\eqref{eq:pip},
\[
\Pi\cdot\Gamma&=q_1\tilde p_1+q_2\tilde p_2\\
&=q_1(p_2-fq_1\nu^\theta)+q_2(-p_1-fq_2\nu^\theta)+\Pi_3\Gamma_3\\
&=q_1p_2-q_2p_1-f\nu^\theta(q_1^2+q_2^2)-\nu^\theta(1-\|q\|^2f)\\
&=(q_1p_2-q_2p_1)-\nu^\theta.
\]
Consequently, the additional conserved momentum (\emph{conserved for
vertical $\nu$ only}) is equivalent to the subcasimir $\Pi\cdot\Gamma$
of~\cite{LeonardMarsden}.
\end{remark}

%
\subsection{Energy-momentum confinement, spectral stability}
\label{sec:stabUWV}
%
%
\begin{theorem}\label{th:KirStabCond}
The axisymmetric relative equilibrium $x_e$, with linear momentum
$P_e$ and spin momentum $S_e$, of the Kirchhoff
model~\eqref{eq:Kirchhoff} is $\kG$-stable by energy-momentum
confinement if
\[[eq:KirEMstab]
mgl>\left(\frac1{M_3}-\frac1{M_1}\right)P_e^2
\]
and spectrally stable if
\[[eq:KirSpecstab]
mgl > \left(\frac1{ M_3}-\frac1{ M_1}\right)P_e^2-\frac{M_1}{4(I_1M_1-m^2l^2)}
 S_e^2.
\]
\end{theorem}
\begin{proof} 
By Proposition~\ref{prop:EStabRE}, it is sufficient to consider the
equilibrium $q=0,p=0$ of the $\RR^3\times \SO2_R$ reduced Kirchhoff
Hamiltonian at the parameter values $\nu^a_3=P_e$ and
$\nu^\theta=S_e$. The relevant Hessian may be obtained
from~\eqref{eq:HamRedVert} by deleting terms of higher than order two,
resulting in
\[[eq:KirHessian]
H=\frac{F_p}2|p+Fq|^2+\frac{F_q}2|q|^2
\]
where, for convenience, we use the notations
\[
q=q_1+iq_2,\quad p=p_1+ip_2,
\quad F=F_l-\frac i2S_e.
\]
By assumption $F_p>0$, so the Hamiltionian~\eqref{eq:KirHessian} is
positive definite when $F_q>0$, a condition which, in view
of~\eqref{eq:Fdef}, is equivalent to~\eqref{eq:KirEMstab}. 

To show~\eqref{eq:KirSpecstab}, note that the linearized equations at
$q=0,p=0$ are obtained from~\eqref{eq:KirHessian} by Hamilton's
equations:
\begin{align*}
&\frac{\d q}{\d t}=\frac{\d q_1}{\d t}+i\frac{\d q_2}{\d t}=
 \frac{\partial H}{\partial p_1}
 +i\frac{\partial H}{\partial p_2}
 =F_pF\,q+F_p\,p,\\
&\frac{\d p}{\d t}=\frac{\d p_1}{\d t}+i\frac{\d p_2}{\d t}=
 -\frac{\partial H}{\partial q_1}
 -i\frac{\partial H}{\partial q_2}
=-(F_p|F|^2+F_q)\,q-F_p\bar F\,p.
\end{align*}
Due to $\SO2$-equivariance, computing the eigenvalues of the
linearization reduces to solving the equation
\[
0=\det\left(\begin{array}{cc}F_pF-\lambda&F_p\\
 -(F_p|F|^2+F_q)&-F_p\bar F-\lambda\end{array}\right) 
 =\lambda^2+iF_pS_e\lambda+F_pF_q.
\]
This has roots
\[[eq:redevals]
\lambda=\frac12\left(-iF_pS_e\pm\sqrt{-F_p^2S_e^2-4F_pF_q}\right),
\]
so spectral stability holds if and only if
\[[eq:KirSpecstab2]
F_p^2S_e^2+4F_pF_q=4F_p\left(mgl-\biggl(\frac1{M_3}-\frac1{M_1}\biggr)P_e^2+\frac{M_1}{4(I_1M_1-m^2l^2)}S_e^2\right)>0,
\]
which is equivalent to~\eqref{eq:KirSpecstab}.\end{proof}

\begin{remark}
Energy-momentum confinement under momentum-preserving perturbations,
and spectral stability, fail coincidentally as predicted in general by
Proposition~\ref{prop:HamHopf}, and as observed in this example
by~\cite{LeonardMarsden}. We can check this by determining the
conditions such that $H+F_p\lambda(q_1p_2-q_2p_1)$ is positive
definite for some $\lambda\in\RR$, where $H$~is the Hessian
in~\eqref{eq:KirHessian}. One computes
\begin{align}
H+\lambda(q_1p_2-q_2p_1)&=
 \frac12\bigl(F_p|p+Fq|^2+2\lambda(q_1p_2-q_2p_1)\bigr)
 +\frac{F_q}2|q|^2\notag\\
&=\frac1{2F_p}|F_p(p+Fq)+i\lambda q|^2+\frac1{2F_p}\bigl(F_pF_q
 +\lambda F_pS_e-\lambda^2\bigr)|q|^2,\label{eq:HlambdaHess}
\end{align}
where we have used
\[
|F_p(p+Fq)+i\lambda q|^2&=F_p^2|p+Fq|^2+\lambda^2|q|^2
 +2\Im\bigl(F_p(p+Fq)(\lambda q)^-\bigr)\\
&=F_p^2|p+Fq|^2+\lambda^2|q|^2+2\lambda F_p\bigl((q_1p_2-q_2p_1)
 +|q|^2\Im F\bigr).
\]
In~\eqref{eq:HlambdaHess}, the coefficient of $|q|^2$ is positive for
some $\lambda$ if and only if $F_pF_q+\lambda F_pS_e-\lambda^2$ has a
real root, i.e.\ if and only if
\[
F_p^2S_e^2-4(F_pF_q)(-1)=F_p^2S_e^2+4F_pF_q>0,
\]
which is exactly the condition for spectral stability as
in~\eqref{eq:KirSpecstab2}.
\end{remark}

%
\subsection{KAM stability}
\label{subsection-KAM-stability}
%
%
In this section we verify that (at vertical $\nu^a$) the equilibrium
$q=0,p=0$ of the $\RR^3\times\SO2$ reduced
Hamiltonian~\eqref{eq:HamRed} satisfies the twist condition of the
Arnold stability theorem (see~\cite{MeyerHall}) in the gap
between~\eqref{eq:KirEMstab} and~\eqref{eq:KirSpecstab}:

\begin{theorem}\label{Th:KAMUWV}
The axisymmetric relative equilibrium $x_e$ of the Kirchhoff
model~\eqref{eq:Kirchhoff} (with linear momentum $P_e$ and spin momentum
$S_e$) is stable if
\[
P_e^2\left(\frac1{M_3}-\frac1{M_1}\right)>mgl>
 P_e^2\left(\frac1{M_3}-\frac1{M_1}\right)
 -\frac{M_1}{4(I_1M_1-m^2l^2)}S_e^2.
\]
\end{theorem}
\noindent For the proof we use Arnold's Theorem, see
e.g.~\cite[Chapter IX.E]{MeyerHall}. This requires us to bring the
Hamiltonian into the normal form
\[[eq:HamArnold]
H=H_2+ H_4\cdots+H_{2N}+H^\dagger
\]
where
\begin{enumerate}
\item 
$H_{2k}$, $1\le k\le 2N$ are homogeneous degree $k$ in the polynomials
$\kI_1=\frac12(q_1^2+p_1^2)$ and $\kI_2=\frac12(q_2^2+p_2^2)$;
\item 
$H^\dagger$ is at least order $2N+1$ in $q_1,q_2,p_1,p_2$;
\item 
$H_2=\omega_1\kI_1-\omega_2\kI_2$, where $\omega_1,\omega_2\neq0$.
\end{enumerate}

\begin{theoremNonumber}[Arnold Stability Theorem]
The origin is stable for the Hamiltonian system~\eqref{eq:HamArnold},
provided for some $k$, $2\leq k\leq N$, the twist condition
$D_{2k}=H_{2k}(\omega_2,\omega_1)\ne0$ is satisfied.
\end{theoremNonumber}

\begin{proof}[Proof of Theorem~\ref{Th:KAMUWV}]
Necessarily $\nu^\theta=S_e\ne0$, or else there is no gap. By choice
of time scale, we can assume that $S_e=1$. Also, $F_q<0$ in view
of~\eqref{eq:Fdef} and~\eqref{eq:KirSpecstab}. The proof requires four steps.

\medskip\noindent{\it 1. Consolidate parameters.} The conditions
$D_{2k}\neq0$ of the Arnold Stability Theorem are algebraically
complicated in the parameters. The number of these conditions that is
expected to be required is one more than the number of free parameters
in the Hamiltonian, so as to have the reasonable expectation that
there will be no value of the free parameters for which all conditions
vanish. To facilitate the symbolic computation of the required normal
form, so that the computation to be tractable, it is necessary to
reduce as much as possible the number of parameters, and to cast the
remaining free parameters so that they appear as simple powers.

First, the parameter $F_l$ of the Hamiltonian~\eqref{eq:HamRedVert} may be
eliminated by the substitution
\[p_1=u_1-\frac{F_lq_1}{\Gamma_3},\qquad p_2=u_2-\frac{F_lq_2}{\Gamma_3}.\]
This is a symplectic transformation 
because it is a shift by the exact form
\[
-F_l\,\d\Gamma_3=\frac{F_lq_1}{\Gamma_3}\,\d q_1
 +\frac{F_lq_2}{\Gamma_3}\,\d q_2.
\]
By a simple direct computation, 
\[
&(p_1+F_l\Gamma_3q_1+ fq_2)^2
 -(q_1p_1+q_2p_2)^2
 +(p_2+F_l\Gamma_3q_2- fq_1)^2\\
&\qquad=(u_1+ fq_2)^2
 -(q_1u_1+q_2u_2)^2+(u_2- fq_1)^2-F_l^2|q|^4
\]
so~\eqref{eq:HamRed} is transformed to
\[
H(q_1,q_2,u_1,u_2)&=
\frac{F_p}2
\bigl((u_1+ fq_2)^2
 -(q_1u_1+q_2u_2)^2+(u_2-fq_1)^2\bigr)\\
&\qquad\mbox{}+\frac{F_q}2\|q\|^2+mgl\left(f-\frac12\right)\|q\|^2,
\]
where, evidently, $F_l$ is absent.

The frequencies of the linearization of $q=0,p=0$ are related to
$F_p$, $F_q$ by~\eqref{eq:redevals}:
\[
f_1=\frac12(F_p+\sqrt{F_p^2+4F_pF_q}),\qquad
f_2=\frac12(F_p-\sqrt{F_p^2+4F_pF_q}),
\]
and these can be inverted to give
\[[eq:fparams]
F_p=f_1+f_2,\qquad F_q=-\frac{f_1f_2}{f_1+f_2}.
\]
Note that $f_1,f_2>0$ and $f_1-f_2=\sqrt{F_p^2+4F_pF_q}>0$ so that
$f_1>f_2>0$. Replacing $mgl$ with $-\mu$, and deleting an inessential
constant, it is sufficient to consider the Hamiltonian
\[[eq:HamToBeNormaled]
H(q_1,q_2,u_1,u_2)&=
\frac{f_1+f_2}2
\bigl((u_1+ fq_2)^2
 -(q_1u_1+q_2u_2)^2+(u_2-fq_1)^2\bigr)\\
&\qquad\mbox{}-\frac{f_1 f_2}{2(f_1+f_2)}\|q\|^2
 +\mu\left(\sqrt{1-\|q\|^2}+\frac12\|q\|^2\right)
\]
under the assumptions that
\[f_1>f_2>0.\]
The number of free parameters is $2$, because an inessential
multiplier of the Hamiltonian can reduce the triple $f_1,f_2,\mu$ by
one. Thus, three twist conditions are expected to be required.

\medskip\noindent{\it 2. Linear normal form.} The linearization at
the origin with respect to the coordinates $(q_1,q_2,u_1,u_2)$ is
\[
\left(\begin{array}{cccc}
 0&\frac12(f_1+f_1)&f_1+f_2&0\\
 -\frac12(f_1+f_2)&0&0&f_1+f_2\\
 -\frac{(f_1-f_2)^2}{4(f_1+f_2)}&0&0&\frac12(f_1+f_2)\\
 0&-\frac{(f_1-f_2)^2}{4(f_1+f_2)}&-\frac12(f_1+f_2)&0
\end{array}\right)
\]
and four linearly independent eigenvectors are
\begin{align*}
&e_1=\left(\begin{array}{cccc}2(f_1+f_2)&0&0&f_2-f_1\end{array}\right)\\
&e_2=\left(\begin{array}{cccc}0&2(f_1+f_2)&f_1-f_2&0\end{array}\right)\\
&e_3=\left(\begin{array}{cccc}0&2(f_1+f_2)&f_2-f_1&0\end{array}\right)\\
&e_4=\left(\begin{array}{cccc}2(f_1+f_2)&0&0&f_1-f_2\end{array}\right)
\end{align*}
with respect to which the symplectic form is canonical with multiplier
$4(f_1-f_2)(f_1+f_2)$, which can be ignored. The coordinates
$(Q_1,P_1,Q_2,P_2)$ defined by
\[
\left(\begin{array}{cccc}q_1&q_2&u_1&u_2\end{array}\right)=
 Q_1e_1+P_1e_2+Q_2e_3+P_2e_4
\]
normalizes the quadratic part of $H$ to
\[
\frac{\omega_1}2(Q_1^2+P_1^2)-\frac{\omega_2}2(Q_2^2+P_2^2),
\]
where
\[
\omega_1=4(f_1-f_2)(f_1+f_2)f_1,\qquad
\omega_2=4(f_1-f_2)(f_1+f_2)f_2.
\]
The $\SO2$ action for vertical $\nu$ explained in
Remark~\ref{rem:redso2mom} is anticlockwise rotations on $(Q_1,P_1)$
and clockwise rotations on $(Q_2,P_2)$. The momentum of this action
can be taken as
\[[eq:linearmom]
\!J=-\frac12(Q_1^2+P_1^2)+\frac12(Q_2^2+P_2^2)=-\kI_1+\kI_2.
\]

\medskip\noindent{\it 3. Nonlinear normal form.}
Since the Hamiltonian $H$ is invariant under an $\SO2$ symmetry, we
choose to seek normalizations within the corresponding class of
invariant functions. Convenient invariants are
\begin{alignat*}{3}
&w_1=\kI_1=\frac12(Q_1^2+P_1^2),&&\qquad w_2=\kI_2=\frac12(Q_2^2+P_2^2),\\
&w_3=\frac1{\sqrt2}(Q_1Q_2-P_1P_2),&&\qquad w_4=\frac1{\sqrt2}(Q_1P_2+Q_2P_1),
\end{alignat*}
and these satisfy the relation
\[[35]
2w_1w_2-w_3^2-w_4^2=0.
\]
The Poisson bracket of the $w_i$ is closed and the matrix
$\{w_i,w_j\}$ is
\[[33]
\{w_i,w_j\}=\left(\begin{array}{cccc}
0&0&w_4&-w_3\\
0&0&w_4&-w_3\\
-w_4&-w_4&0&-w_1-w_2\\
w_3&w_3&w_1+w_2&0
\end{array}\right)
\]

Since $H$ is invariant it can be expanded in a Taylor
series in $w_1,w_2,w_3,w_4$ as
\[
H=H_2+H_4+\cdots+H_{2N}+H^\dagger
\]
where
\[
H_2=\omega_1\kI_1-\omega_2\kI_2=\omega_1w_1-\omega_2w_2
\] 
and the $H_{2i}$ are degree~$i$ homogeneous polynomials in
$w_1,w_2,w_3,w_4$. The normalization process is formally the same in
the space of invariants as it is in the space of functions of
$q_1,p_1,q_2,p_2$. See~\cite[Chapter VII.C]{MeyerHall}. The normal
form is achieved at degree~$k\ge2$ successively in $k$. After
normalizing the terms of degree smaller than or equal to $k$, at the
beginning of step $k$, we start with the Hamiltonian $H_{2k}$ from the
previous step. The Hamiltonian $H_{2k}$ is in normal form up to order
$k$, i.e.\ it is the sum of homogeneous degree $i\le k$ polynomials in
$w_1,w_2$ and homogeneous degree $i>k$ polynomials in
$w_1,w_2,w_3,w_4$. We now look for a homogeneous polynomial $G$ of
degree $k+1$ in $w_1,w_2,w_3,w_4$ such that the time~$1$ flow of $G$
transforms $H$ into a function
\[
\tilde H=H+\{G,H\}+\frac1{2!}\{G,\{G,H\}\}+\cdots,
\]
which is in normal form at degree $k+1$. The degree $k+1$ term of
$\tilde H$ is $H_{2(k+1)}+\{G,H_2\}$ because
\[
\deg\{G,H_{2i}\}=\deg\D G+\deg\D H_{2i}+1=k+(i-1)+1=k+i,
\]
so $\deg\{G,H_{2i}\}=k+1$ only for $i=1$. The coefficients of $G$ are
adjusted so that 
\[[eq:homological]
\tilde H_{2k}=H_{2k}+\{G,H_2\}
\] 
is a function only of $w_1$ and $w_2$, after all powers of $w_4$
greater than $1$ are eliminated using~\eqref{35}. The Poisson brackets
here are computed using~\eqref{33}. Then $H$ is replaced by $\tilde H$
and the computation proceeds to step~$k+1$.

At step~$k$, there always exists a $G$ such that $\tilde H_{2k}$ is
normalized. Indeed, $\{\cdot,H_2\}$ maps the finite dimensional
vector space $\PP_{k+1}$ of homogeneous degree~$k+1$ polynomials to
itself. The space $\PP_{k+1}$ can be regarded as an inner product space by
\[
\langle f,g\rangle=\int_{B_{12}\times B_{34}}fg
\]
where $B_{12},B_{34}$ are the unit disks in the $w_1,w_2$ and
$w_3,w_4$ variables. The linear map $\{\cdot,H_2\}$ is antisymmetric
with respect to this inner product because
\[
\langle f,\{g,H_2\}\rangle+\langle \{f,H_2\},g\rangle&=
\int_{B_{12}\times B_{34}}\{fg,H_2\}\\
&=\int_{B_{12}\times B_{34}}(\omega_1-\omega_2)
 \left(-w_4\frac{\partial(fg)}{\partial w_3}
 +w_3\frac{\partial(fg)}{\partial w_4}\right)\\
&=-(\omega_1-\omega_2)\int_{B_{12}}\ointctrclockwise_{\partial B_{34}}
 fg(w_3\,\d w_3+w_4\,\d w_4)\\
&=0
\]
because the vector field in the line integral is always orthogonal to
its path. Consequently, the image of $\{\cdot,H_2\}$ is orthogonal to
$\ker\{\cdot,H_2\}$ and $G$ in~\eqref{eq:homological} can be chosen
so that $\tilde H_{2k}\in\ker\{\cdot,H_2\}$. Moreover, $f\in\PP_{k+1}$
is in $\ker\{\cdot,H_2\}$ if and only if
\[
(\omega_1-\omega_2)\left(-w_4\frac{\partial f}{\partial w_3}
 +w_3\frac{\partial f}{\partial w_4}\right)=0,
\]
from which $f=\tilde f(w_1,w_2,w_3^2+w_4^2)$, since
$\omega_1\ne\omega_2$. But in view of the relation~\eqref{35}, this
implies that $f$ is a function of $w_1,w_2$, as required. We note that
only the nonresonance condition $\omega_1\ne\omega_2$ is required and
the normalization could in theory be achieved at any order.

The normal form cannot be hand-computed, but can be computed using a
symbolic manipulator. The result up to and including $H_8$ (the
notation $1\leftrightarrow2$ means the preceeding fragment with
$w_1$ and $w_2$ exchanged) is
\begin{align*}
&H_4=8(f_1+f_2)^3
 \Bigl(A_{20}w_1^2+\frac12A_{11}w_1w_2+1\leftrightarrow2\Bigr),\\
&H_6=-\frac{32(f_1+f_2)^5}{(f_1-f_2)^2}
 \Bigl(A_{30}w_1^4+A_{31}w_1^2w_2+1\leftrightarrow2\Bigr),\\
&H_8=-\frac{64(f_1+f_2)^7}{(f_1-f_2)^4}\Bigl(A_{40}w_1^4
 +A_{41}w_1^3w_2+\frac12A_{42}w_1^2w_2^2+1\leftrightarrow2\Bigr),\\\\
&A_{20}=(-f_1-f_2)\mu+(f_1+f_2)f_2 ,\\
&A_{21}=(-4 f_1-4 f_2)\mu +8 f_1f_2 ,\\
&A_{30}=2(f_1+f_2)^{2}\mu^{2}-2 (f_1+f_2)(3 f_1+f_2)\mu f_2+4( f_1+f_2)f_1f_2^{2} ,\\
&A_{31}=15(f_1+f_2)^{2}\mu^{2}-(f_1+f_2)(5f_1^{2}+44 f_1f_2 +11f_2^{2})\mu\ebrk+(5f_1^{2}+38 f_1f_2+17f_2^{2}) f_1f_2 ,\\
&A_{32}=15(f_1+f_2)^{2}\mu^{2}-(f_1+f_2)(11f_1^{2}+44 f_1 f_2+5f_2^{2})\mu\ebrk+(17f_1^{2}+38 f_1f_2+5f_2^{2}) f_1f_2 ,\\
&A_{40}=-16(f_1+f_2)^{3}\mu^{3}+2 (f_1^{2}+36 f_1f_2+11f_2^{ 2})(f_1+f_2)^{2}\mu^{2}-2 (f_1+f_2)(2f_1^{3}\ebrk+55 f_1^{2}f_2+36 f_1f_2^{2}+3 f_2^{3})\mu f_2\ebrk+2(f_1+f_2) (f_1^{2}+26 f_1f_2+5f_2^{2 }) f_1f_2^{2} ,\\
&A_{41}=-182(f_1+f_2)^{3}\mu^{3}+26(3f_1^{2}+31 f_1f_2+8f_2^{2})(f_1+f_2)^{2}\mu^{2}\ebrk-2(13 f_1+f_2)(f_1+f_2)(8f_1^{2}+49 f_1f_2+21f_2^{2})\mu f_2\ebrk+2(65f_1^{3}+ 367f_1^{2}f_2+267 f_1f_2^{2}+29 f_2^{3}) f_1f_2^{2} ,\\
&A_{42}=-354(f_1+f_2)^{3}\mu^{3}+3(95f_1^{2}+518 f_1f_2+95f_2^{2})(f_1+f_2)^{2}\mu^{2}-3(f_1+f_2)\ebrk(9f_1^{4}+274f_1^{3}f_2+850f_1^{2}f_2^{2}+274 f_1f_2^{3} +9f_2^{4})\mu +3(9f_1^{4}+204f_1^{3}f_2\ebrk+518f_1^{2}f_2^{2}+204 f_1f_2^{3} +9f_2^{4}) f_1f_2.
\end{align*}

\medskip\noindent{\it 4. Twist condition} The twist condition in the
Arnold stability theorem is $D_{2k}=H_{2k}(\omega_2,\omega_1)\ne0$ for
some $k\ge 2$. From the normal form, the first three $D_{2k}$ are
\begin{align*}
&D_4=128(f_1-f_2)^{2}(f_1+f_2)^{5}(-(f_1+f_2)(f_1^{2}+4 f_1f_2+f_2^{2})\mu\ebrk+(f_1^{2} +10 f_1f_2+f_2^{2})f_1f_2),\\[1ex]
&D_6=-2048(f_1+f_2)^{9}((2f_1^{2}+13 f_1f_2+2f_2^{2})(f_1+f_2)^{2}\mu^{2}-(f_1+f_2)(11f_1^{2}\ebrk+46 f_1f_2+11f_2^{2})\mu f_1f_2+(9f_1^{2}+ 50 f_1f_2+9f_2^{2})f_1^{2 }f_2^{2})(f_1-f_2),\\[1ex]
&D_8=16384(f_1+f_2)^{11}(-2(8f_1^{4}+91f_1^{3}f_2+177f_1^{2}f_2^{2}+91 f_1f_2^{3}+ 8f_2^{4})(f_1+f_2)^{3}\mu^{3}\ebrk+(2f_1^{6}+150f_1^{5}f_2+1113f_1^{4}f_2^{2}+1970f_1^{3}f_2^{3}+ 1113f_1^{2}f_2^{4}\ebrk+150 f_1f_2^{5}+2f_2^{6})(f_1+f_2)^{2}\mu^{2}-(f_1+f_2)(4f_1^{6}+345f_1^{5}f_2+ 2226f_1^{4}f_2^{2}\ebrk+3850f_1^{3}f_2^{3}+2226f_1^{2 }f_2^{4}+345 f_1f_2^{5}+4f_2^{6})\mu f_1f_2+(2f_1^{6 }+211f_1^{5}f_2\ebrk+1466f_1^{4}f_2^{2}+2642f_1^{3}f_2^{3}+1466f_1^{2}f_2^{4}+ 211 f_1f_2^{5}+2f_2^{6})f_1^{2}f_2^{2}).
\end{align*}
We assume $D_4,D_6,D_8$ vanish and argue by contradiction. The
expressions are homogeneous in $f_1,f_2$, so one can assume $f_2=1$.
Since $D_4$ is linear in $\mu$, if the coefficient of $\mu$ vanishes,
then the assumption $D_4=0$ implies the constant term in $\mu$ also
vanishes, i.e.\
\[
-(f_1+1)(1+4 f_1+f_1^{2})=0,\quad\mbox{and}\quad f_1(f_1^{2}+10 f_1+1)=0.
\]
But these two polynomials in $f_1$ cannot simultaneously vanish
because their gcd is~1, as can be computed with the Euclidean
algorithm. So the coefficient of $\mu$ in $D_4$ does not
vanish. Solving $D_4$ for $\mu$ and substituting into $D_6$ and $D_8$
results in
\begin{align*} 
&0=10240(f_1+1)^{9}f_1^{3}(f_1+5)(5f_1+1)(f_1-1)^{3} \\
&0=-245760(f_1+1)^{11}f_1^{4}(21f_1^{4}+172f_1^{3}+334f_1^{2}+172f_1+21)(f_1-1)^{4}
\end{align*}
But these cannot simultaneously vanish because their gcd is
$10240(f_1+1)^{9}f_1^{3}(f_1-1)^{3}$, which is nonzero in view of
$f_1>f_2=1$.
\end{proof}

\begin{remark}
The computer generated normal form used in Theorem~\ref{Th:KAMUWV}
requires an independent check, and one possible check follows.  The
reduction by $\SO2$ of the $\SO2$-invariant
Hamiltonian~\eqref{eq:HamToBeNormaled} gives a one parameter family of
2-dimensional Hamiltonian systems parametrized by the $\SO2$ angular
momentum. All trajectories of these reduced Hamiltonian systems near
$q=0,u=0$ are periodic. The periods of these orbits depend only on
the energy and the $\SO2$~momentum, and may be numerically
computed. On the other hand, the successive normal form Hamiltonians
\[[eq:nfs]
&H^{\text{nf}}_4=H_2+H_4,\\
&H^{\text{nf}}_6=H_2+H_4+H_6,\\
&H^{\text{nf}}_8=H_2+H_4+H_6+H_8,
\]
are functions of $w_1$ and $w_2$. Using the Poisson
bracket~\eqref{33}, the differential equations for $w_i$ are
\[
\frac{\d w_1}{\d t}=\frac{\d w_2}{\d t}=0,\quad
\frac{\d w_3}{\d t}=-\left(\frac{\partial H^{\text{nf}}_{2i}}{\partial w_1}
+\frac{\partial H^{\text{nf}}_{2i}}{\partial w_2}\right)w_4,\quad
\frac{\d w_4}{\d t}=\left(\frac{\partial H^{\text{nf}}_{2i}}{\partial w_1}
+\frac{\partial H^{\text{nf}}_{2i}}{\partial w_2}\right)w_3.
\]
These differential equations are the $\SO2$ reduction of the
Hamiltonian system~\eqref{eq:HamToBeNormaled}, and the orbits have periods
\[[eq:nfpers]
T^{\text{nf}}_{2i}=\left|\frac{\partial H^{\text{nf}}_{2i}}{\partial w_1}
+\frac{\partial H^{\text{nf}}_{2i}}{\partial w_2}\right|,
\]
which also depend only on the energy and the $\SO2$~momentum. For
energy and momentum obtained from initial conditions $\epsilon
q^0,\epsilon p^0$, the periods of $H^{\text{nf}}_{2i}$, computed
directly from~\eqref{eq:nfpers}, can be compared with numerically
computed periods of the Hamiltonian~\eqref{eq:HamToBeNormaled}. If the
normal form is correct, then the difference of these must fall as
$\epsilon^{2i}$, because the period from $H^{\text{nf}}_{2i}$ is
accurate to order $i-1$ in $w$, and therefore its difference falls as
order $i$, which is order $2i$ in $\epsilon$. If the normal form is
not correct, then it is not likely that the anharmonic periods it
predicts will agree at the proper order with the periods of its
non-normalized precursor.

\begin{table}[t]
\hspace*{-.15in}\rlap{\begin{tabular}{|c||c|cc|cc|cc|}
\hline
$10^4\epsilon$&$T(\epsilon)/T_0$&\multicolumn{2}{c|}{$2i=4$}&\multicolumn{2}{c|}{$2i=6$}&\multicolumn{2}{c|}{$2i=8$}\\
&&$T^{\text{nf}}_4(\epsilon)/T_0$&$r_{4,\epsilon}$&$T^{\text{nf}}_6(\epsilon)/T_0$&$r_{6,\epsilon}$&$T^{\text{nf}}_8(\epsilon)/T_0$&$r_{8,\epsilon}$
\\
\hline\hline
1&.9995410553688&.99954076 &4.0&.99954105570 & 6.0&.9995410553684&8.2\\
2&.9981715477722&.99816682 &4.0&.99817156906 &6.0&.9981715476563& 8.0\\
4&.9928005386538&.99272712 &3.3&.99280185244 &5.8& .9928005101406&7.8\\
8&.9728669887964&.97182458 &3.5& .97294031478 &5.4&.9728607222126&7.2\\
16&.9108473444693&.89972618 && .91385268217 &&.9098792303872&\\
\hline
\end{tabular}}
\caption{\label{nf-ok-table}\it Comparison, at the same
energy-momentum values, of numerically computed periods, and periods
computed from the normal forms $H^{\text{\rm nf}}_{2i}$. The correct
normal form is indicated because the values of $r$ in the columns are
nearly the corresponding values of $2i$.}
\end{table}
%
%
%
%

We have implemented this check for the data
\[
&f_1=\frac{\sqrt5}2,\quad f_2=1,\quad\mu=\frac12,
\quad q^0=(0.5,1.0),\quad p^0=(-0.75,0),\\
&\epsilon=0.0001,\;0.0002,\;0.0004,\;0.0008,\;0.0016.
\]
The choice of $f_1$ is convenient because then $4(f_1-f_2)(f_1+f_2)=1$
and the symplectic multiplier of Theorem~\ref{Th:KAMUWV} does not have
to be accounted for. In Table~\ref{nf-ok-table} are the numerically
computed periods $T(\epsilon)$, and the periods
$T^{\text{nf}}_{2i}(\epsilon)$ computed using the normal forms, as
ratios of the period
\[
T_0=\frac{2\pi}{f_1-f_2}=\frac{2\pi}{\frac{\sqrt5}2-1},
\]
of the linearization. Also tabulated are the values
called~$r_{2i,\epsilon}$, which are the base two logarithms of the
successive (as $\epsilon$ is successively doubled) ratios of the
differences between the periods $T^{\text{nf}}_{2i}(\epsilon)$ and
$T(\epsilon)$, i.e.
\[
r_{2i,\epsilon}=\log_2\left|\frac{T^{\text{nf}}_{2i}(2\epsilon)-T(2\epsilon)}
 {T^{\text{nf}}_{2i}(\epsilon)-T(\epsilon)}\right|.
\]
In the column corresponding to $H^{\text{nf}}_{2i}$, there is
agreement between $r_{2i,\epsilon}$, which is the observed order that the
difference falls as, and $2i$, which is the predicted order. We
conclude that there is numerical evidence that the normal forms
$H^{\text{nf}}_{2i}$ are correct, because they predict the correct
anharmonic period with an error of order~$2i$.
\end{remark}

%
\section{Numerics}
\label{sec:numerics}
%
%
{\ParagraphLadybird{bug8.jpg}{.35}
In this section we simulate small dissipation directly in the
$\RR^3\times\SO2_R$ reduction of the Kirchhoff model, as computed in
Proposition~\ref{prop:SO2Rred}. We verify that, after addition of
small dissipation,
\begin{enumerate}
\item 
stability is maintained for the axisymmetric relative equilibria $x_e$
which are in the EM-region determined
by~\eqref{eq:KirEMstab};
\item
stability is destroyed for the axisymmetric relative equilibria $x_e$
which are in the gap determined by~\eqref{eq:KirEMstab}
and~\eqref{eq:KirSpecstab}.
\end{enumerate}}
\noindent This confirms that the stability of the system in the gap is
qualitatively less robust than its stability in the EM-region, and
therefore confirms that \emph{energy-momentum confinement fails in the
gap, with physically significant consequences}. The approach of
simulating the $\RR^3\times\SO2_R$ reduced system, rather than the
full Kirchhoff model~\eqref{eq:LKirchhoff}, is consistent with our
development in Sections~\ref{sec:EuclAxiRE} and~\ref{sec:EuclStab}, and
is convenient for the simulation of energy dissipation which preserves
the $\RR^3\times\SO2_R$ momentum and symmetry (see
Remark~\ref{rm:diss}).

%
\subsection{Splitting method}
\label{sec:splitting}
%
%
The Hamiltonian~\eqref{eq:HamRed} may be split as
\[
H=H_0+H_1+H_2+H_3
\]
where
\begin{align*}
&H_0=H(q_1,q_2,0,0),\\
&H_1=\frac{F_p}2(p_1^2+p_2^2-2q_2p_2q_1p_1-q_1^2p_1^2-q_2^2p_2^2),\\
&H_2=F_pF_l(q_1p_1+q_2p_2)\sqrt{1-q_1^2-q_2^2},\\
&H_3=-\frac{F_p\nu^\theta(p_2q_1-p_1q_2)}{1+\sqrt{1-q_1^2-q_2^2}}.
\end{align*}
Only $H_0$ depends on $\nu_a$, and all but $H_0$ are $\SO2$ invariant
under the diagonal action of $\SO2$ on $(q_1,q_2),(p_1,p_2)$, whereas
$H_0$ has this invariance only for vertical $\nu$ (see
Remark~\ref{rem:redso2mom}).

There are exact formulas for the flows of each $H_i$, as follows:
\begin{enumerate}
\item $H_0$ is a function of $q$ only and hence its flow is 
$p\mapsto p-t\nabla H_0$.
\item The flow of $H_1$ corresponds to the motion of a particle of
mass $1/F_p$ freely moving on a sphere of radius~$1$. Indeed, for the
top half of the sphere with coordinates
$x=q_1\!\e_1+q_2\!\e_1+\sqrt{1-q_1^2-q_2^2}\,\!\e_3$, the kinetic energy
metric of the particle motion, and its inverse, are respectively
\[
\frac1{F_p\Gamma^2}\left(\begin{array}{cc}1-q_2^2&q_1q_2
 \\q_1q_2&1-q_1^2\end{array}\right),\qquad
F_p\left(\begin{array}{cc}1-q_1^2&-q_1q_2\\-q_1q_2&1-q_2^2
 \end{array}\right),
\]
the second of which is twice the matrix corresponding to the quadratic
form $H_1$.
\item Since $H_2$ is linear in $p$, and $p_1,p_2$ are separated, the
differential equations for $q_1,q_2$ close. As is easily verified,
$q_1/q_2$ is conserved for these equations and so $q_1,q_2$ evolve
along radial lines. The radial equation can be separated and
integrated, and the evolution of $p_1,p_2$ follows that from
conservation of energy and momentum.
\item As is easily verified, $q_1^2+q_2^2$ is conserved for the flow
of $H_3$. Using this, the differential equations corresponding to $H_3$
become linear and may be integrated.
\end{enumerate}

For a dissipation vector field, one could use a positive multiple of
the negative gradient in $p_1,p_2$ of the reduced
Hamiltonian~\eqref{eq:HamRed}, i.e.\ the vector field
\[[eq:dvf1]
&\left(-(1-q_1^2)p_1+q_1q_2p_2
 -\frac{ml\nu^a_3\Gamma_3}{M_1}q_1
 -\frac{\nu^\theta}{1+\Gamma_3}q_2\right)\frac\partial{\partial p_1}\\
&\qquad\mbox{}+\left(q_1q_2p_1-(1-q_2^2)p_2
 -\frac{ml\nu^a_3\Gamma_3}{M_1}q_2
 +\frac{\nu^\theta}{1+\Gamma_3}q_1\right)\frac\partial{\partial p_2}.\\
\]
This does not preserve the momentum $q_1p_2-p_1q_2$ which
occurs for vertical $\nu$ (see Remark~\ref{rem:redso2mom}). But the
direction of~\eqref{eq:dvf1} relative to the $\ker\D H$ hyperplane
(for $q_1,q_2$ constant) is unaltered by projection: any projection
of~\eqref{eq:dvf1} onto any subspace will still dissipate energy. So
we project onto the line tangent to the level set of the function
$q_1p_2-p_1q_2$, i.e.\ onto the vector $(q_1,q_2)$. Since the inner
product of ~\eqref{eq:dvf1} and $(q_1,q_2)$ is
\[
&(-1+q_1^2+q_2^2)q_1p_1+(-1+q_1^2+q_2^2)q_2p_2-\frac{ml\nu^a_3\Gamma_3}{M_1}
 (q_1^2+q_2^2)\\
&\qquad\mbox{}=-\Gamma_3^2(q_1p_1+q_2p_2)
 -\frac{ml\nu^a_3\Gamma_3}{M_1}(q_1^2+q_2^2),
\]
we use for the dissipative perturbation vector field $\epsilon R$,
where $\epsilon$ is small and positive, and
\[[eq:dvf2]
R&=\left(-(q_1p_1+q_2p_2)\Gamma_3-F_l(q_1^2+q_2^2)
 \right)q_1\frac\partial{\partial p_1}\\
&\qquad\mbox{}+\left(-(q_1p_1+q_2p_2)\Gamma_3-F_l(q_1^2+q_2^2)\right)q_2\frac\partial{\partial p_2}.
\]
The differential equations corresponding to this vector field are
linear and easily solved exactly.

The full system, dissipation included, is given by the vector field
\[[eq:simulatedthis]
\epsilon R+\sum_{i,s}\left(\frac{\partial H_s}{\partial p_i}\frac\partial{\partial q_i}
-\frac{\partial H_s}{\partial q_i}\frac\partial{\partial p_i}\right).
\]
Letting the flows of the Hamiltonians $H_s$, $0\le s\le3$, be
$F^{H_s}_t$, and the flow of the dissipation~$\epsilon R$ be $G_t$,
the concatenation
\[
G_{\Delta t/2}F^{H_0}_{\Delta t/2}F^{H_3}_{\Delta t/2}F^{H_2}_{\Delta t/2}F^{H_1}_{\Delta t}F^{H_2}_{\Delta t/2}F^{H_3}_{\Delta t/2}F^{H_0}_{\Delta t/2}G_{\Delta t/2}
\]
is a second order one step method, which we use to simulate the
system. At vertical $\nu$, this method preserves the conserved
quantity $q_1p_2-q_2p_1$ because every one of its steps does that
separately. The method is symplectic if $\epsilon=0$.

%
\subsection{Simulations}
\label{sec:simulations}
%
%
The simulations that we report in this section are done at parameter
values
\[[eq:numparams]
&I_1=4,\quad M_1=1,\quad M_3=\frac12,\quad m=1,\quad l=1,\quad g=1,\quad S_e=6,\\
&\nu^a_3=P_e,\quad \nu^\theta=S_e,
\]
i.e.\ units are such that $g=1$. The initial conditions are chosen so
that
\[
q_2=0,\quad p_1=0,\quad p_2=0,\quad \nu^a_2=0.
\]
The perturbation from the equilibrium is achieved by choosing small
nonzero $q_1$ and $\nu^a_1$. From~\eqref{th:KirStabCond},
energy-momentum confinement occurs for $P_e^2<C_1$ and spectral
stability occurs for $P_e^2<C_2$, where
\[
C_1=\frac{M_1M_3}{M_1-M_3}\,mgl,\quad
C_2=\frac{M_1M_3}{M_1-M_3}\left(mgl+\frac{M_1}{4(I_1M_1-m^2l^2)}S_e^2\right).
\]
For the parameter values \eqref{eq:numparams}, $C_1=1$ and $C_2=4$, so
\begin{alignat*}{3}
&2< P_e\colon&&\qquad\text{spectral instability;}\\
&1< P_e<2\colon&&\qquad\text{gap; spectral stability;}\\
&0\le P_e<1\colon&&\qquad\text{EM-region; stability.}
\end{alignat*}

%
\subsubsection{Stability}
\label{sec:numericStablity}
%
%
On the left,
\begin{figure}[t]\setlength{\unitlength}{1in}
\renewcommand{\-}[1]{\mskip-\dbt#1mu} 
\begin{picture}(0,3.25)(-.125,0)
\put(.1,1.85){\makebox(0,0)[bl]{\includegraphics{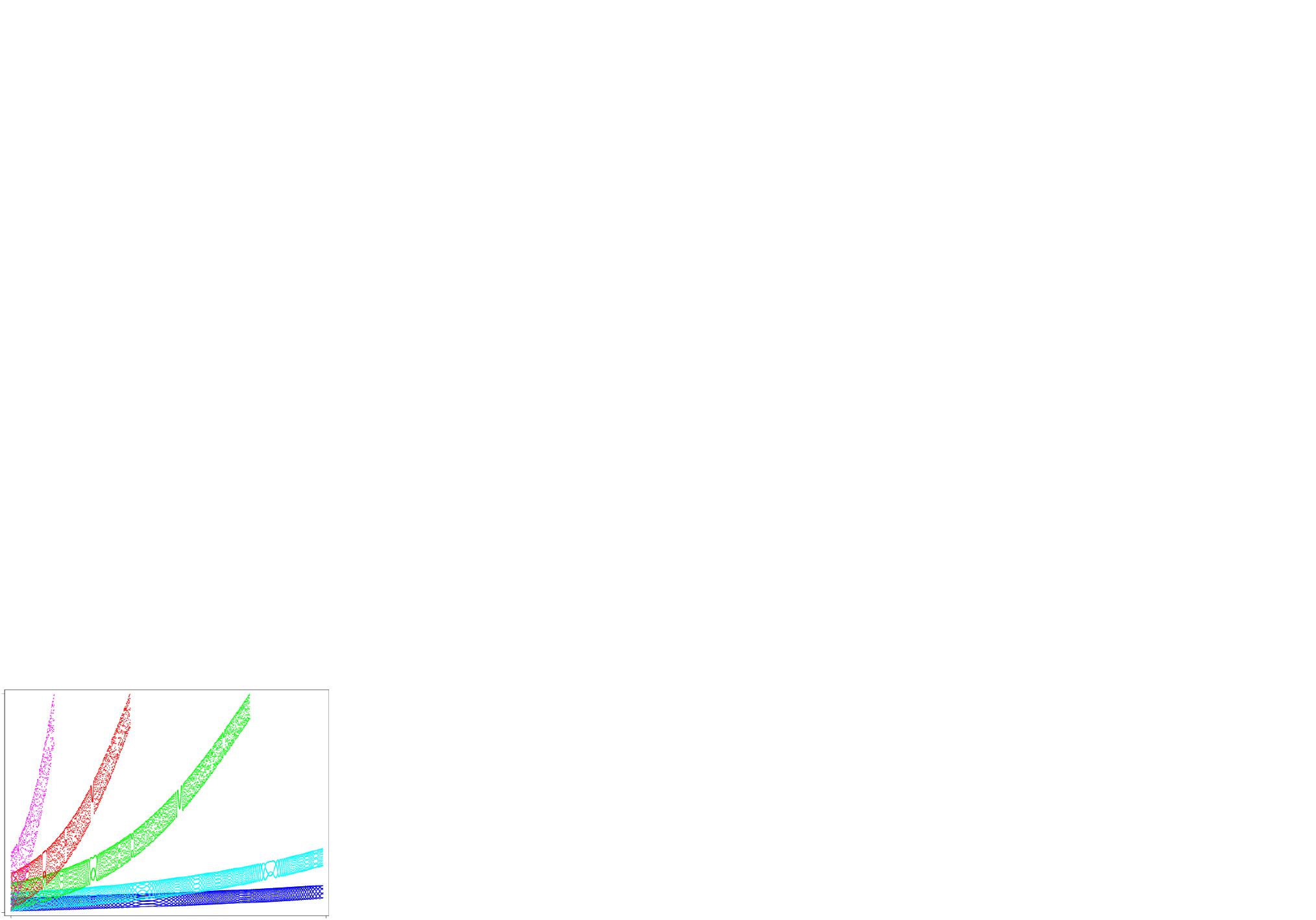}}}
\put(4.9,1.85){\makebox(0,0)[br]{\includegraphics{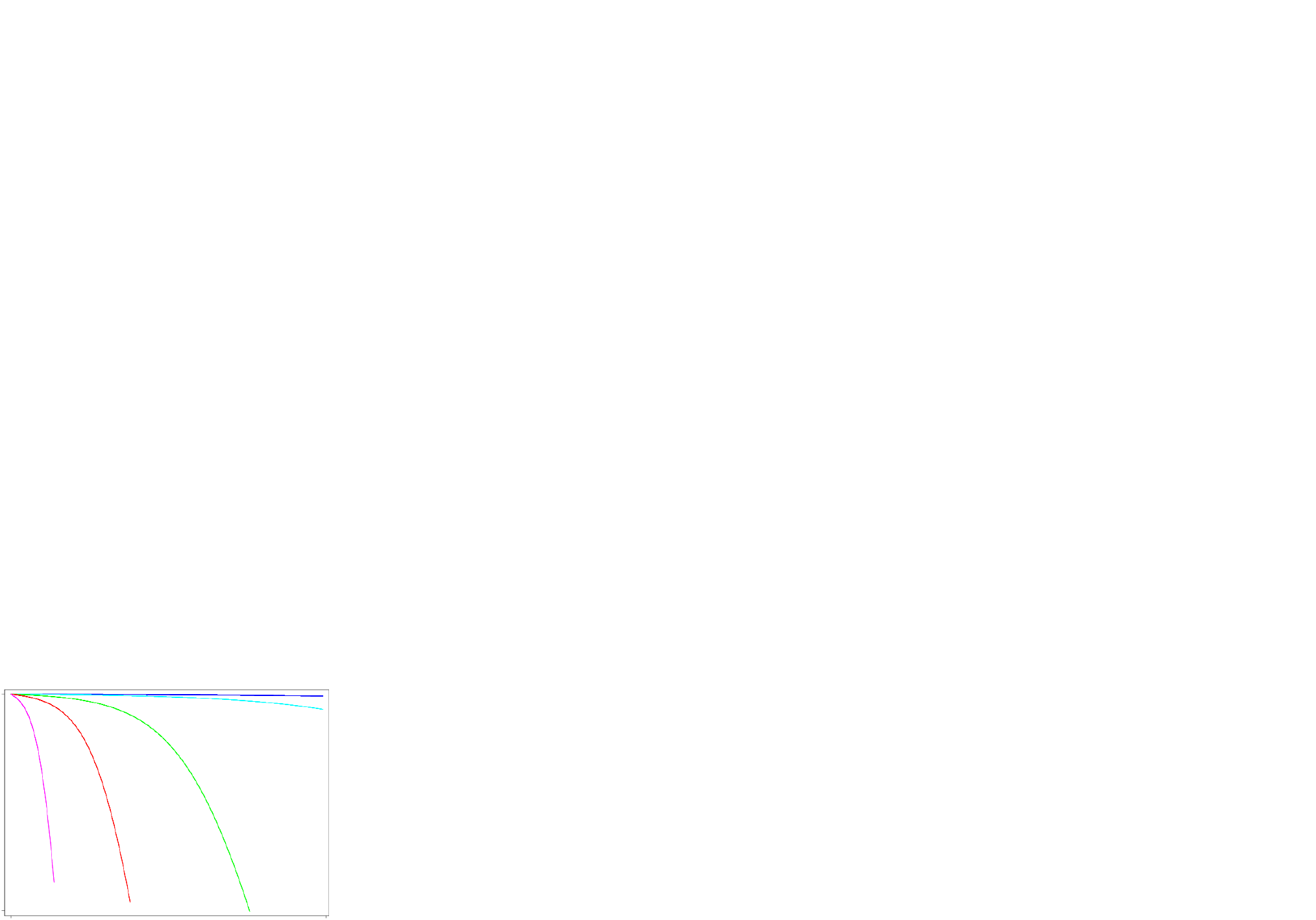}}}
\put(.1,.1){\makebox(0,0)[bl]{\includegraphics{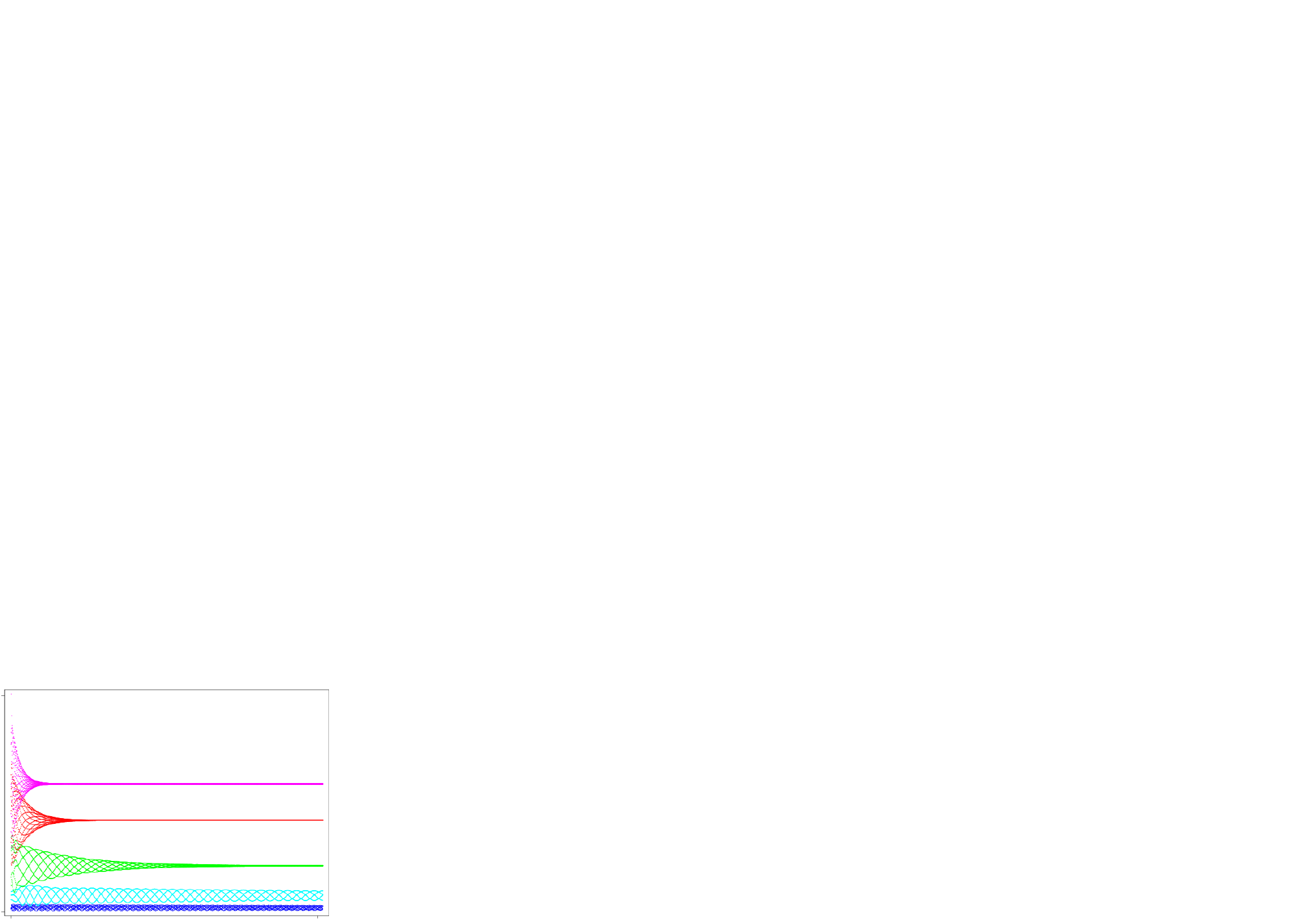}}}
\put(4.9,.1){\makebox(0,0)[br]{\includegraphics{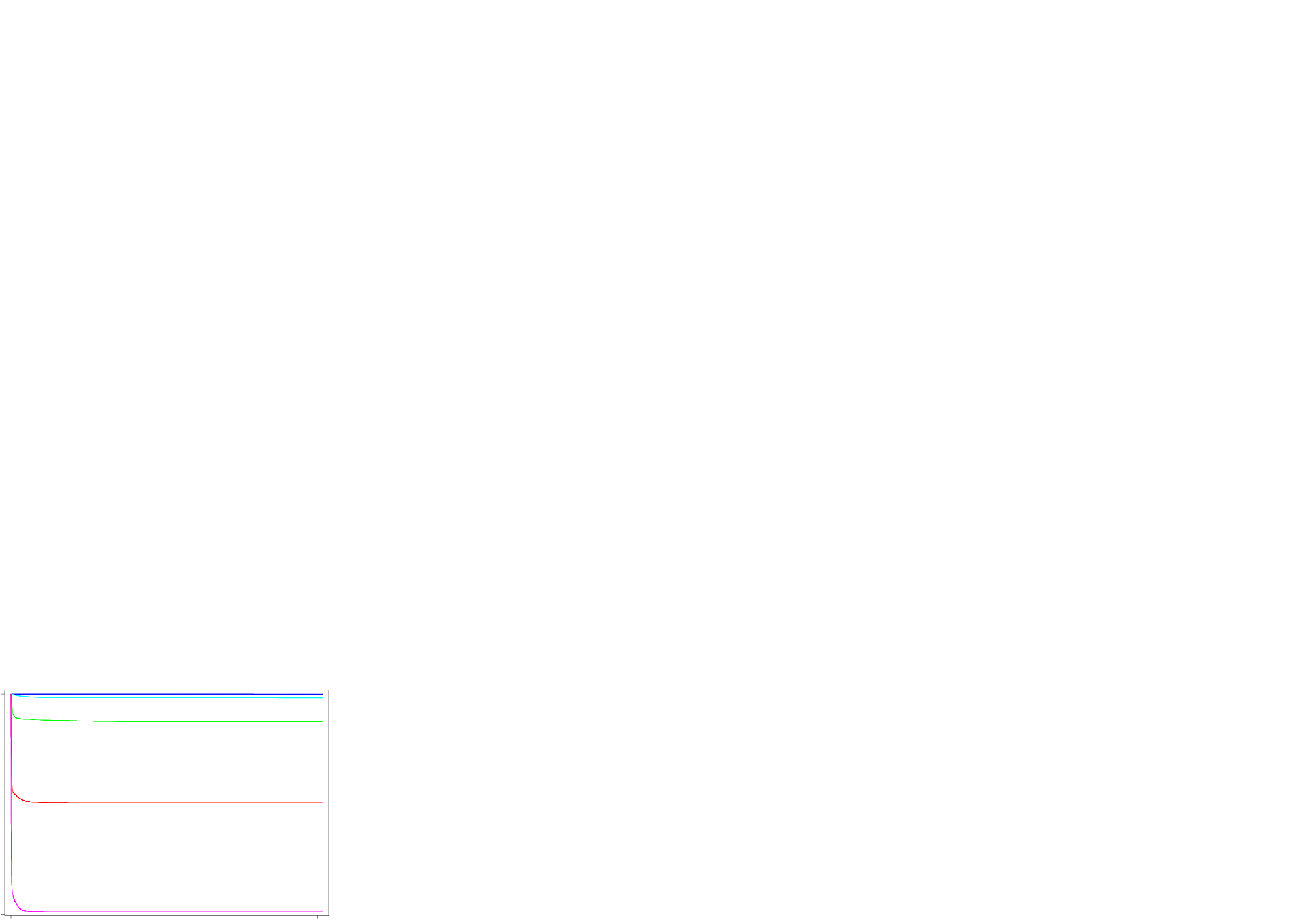}}}
\OrnamentLadybird{-.25}{1}{.3}{-145}
\OrnamentLadybird{2.4}{2}{.3}{36}
\OrnamentLadybird{4}{1}{.3}{-180}
\put(.1725,.075){\makebox(0,0)[tc]{\footnotesize$0$}}
\put(1.9125,.075){\makebox(0,0)[tc]{\footnotesize$3.5\times10^6$}}
\put(1.1,.025){\makebox(0,0)[tc]{\footnotesize$t$}}
\put(.075,.15){\makebox(0,0)[cr]{\footnotesize$0$}}
\put(.075,1.48){\makebox(0,0)[cr]{\footnotesize$.11$}}
\put(0.025,.8){\makebox(0,0)[cr]{\footnotesize$r$}}
\put(2.935,.075){\makebox(0,0)[tc]{\footnotesize$0$}}
\put(4.6875,.075){\makebox(0,0)[tc]{\footnotesize$3.5\times10^6$}}
\put(3.885,.025){\makebox(0,0)[tc]{\footnotesize$t$}}
\put(2.85,.145){\makebox(0,0)[cr]{\footnotesize$-.004$}}
\put(2.85,1.485){\makebox(0,0)[cr]{\footnotesize$0$}}
\put(2.80,.8){\makebox(0,0)[cr]{\footnotesize$\Delta\-2 H$}}
\put(.1725,1.825){\makebox(0,0)[tc]{\footnotesize$0$}}
\put(1.9,1.825){\makebox(0,0)[tc]{\footnotesize$1.08\times10^6$}}
\put(1.1,1.775){\makebox(0,0)[tc]{\footnotesize$t$}}
\put(.075,1.9){\makebox(0,0)[cr]{\footnotesize$0$}}
\put(.075,3.25){\makebox(0,0)[cr]{\footnotesize$.5$}}
\put(0.025,2.55){\makebox(0,0)[cr]{\footnotesize$r$}}
\put(2.935,1.825){\makebox(0,0)[tc]{\footnotesize$0$}}
\put(4.6460,1.825){\makebox(0,0)[tc]{\footnotesize$1.08\times10^6$}}
\put(3.885,1.775){\makebox(0,0)[tc]{\footnotesize$t$}}
\put(2.85,1,91){\makebox(0,0)[cr]{\footnotesize$-.12$}}
\put(2.85,3.24){\makebox(0,0)[cr]{\footnotesize$0$}}
\put(2.80,2.55){\makebox(0,0)[cr]{\footnotesize$\Delta\-2 H$}}
\end{picture}
\caption{\label{numeric-stabustab}\it Left top: instability in the
gap. Left bottom: stability in the EM-region. Dissipation
is added in both the top and bottom, and the energy decay is shown in
the graphs at the right.}
\end{figure}
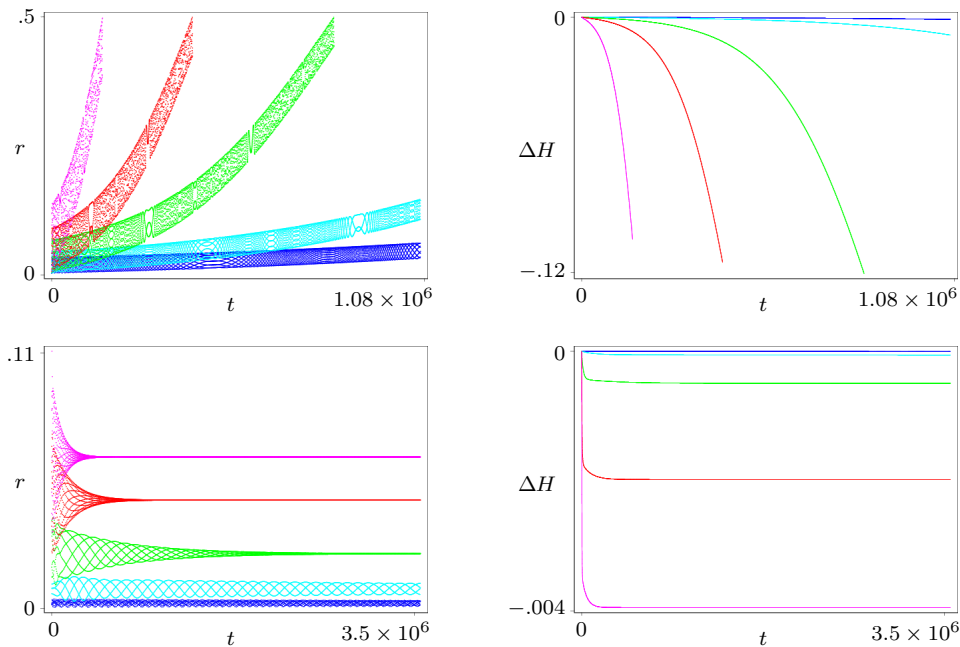
Figure~\ref{numeric-stabustab} shows a regular time
sampling of the radius $r=\sqrt{q_1^2+q_2^2}$ vs. time. The top
and bottom graphs correspond respectively to $P_e=1.5$ (which is
inside the gap) and $P_e=.5$ (which is inside the
EM-region). The graphs have been obtained
from
\begin{alignat*}{7}
&\epsilon=.05,&&\quad q_1=0.05/r,&&\quad \nu^a_1=0.01/k,
 &&\quad k=1,\;1.5,\;2,\;3,\;4,\\
&\epsilon=.1,&&\quad q_1=0.05/r,&&\quad \nu^a_1=0.01/k,
 &&\quad k=\sqrt2,\;(\sqrt2)^2,\;(\sqrt2)^3,\;(\sqrt2)^6,\;(\sqrt2)^{10}.
\end{alignat*} 
and for $3\times10^5$ and $1.2\times10^6$ periods, respectively.
Here, ``period'' means the smallest period of the normal modes of the
linearization at the equilibrium $q=p=0$, i.e.\ $2\pi/\omega$ where
$\omega$ is the largest of the imaginary parts of the eigenvalues
computed from~\eqref{eq:redevals}. The top and bottom of
Figure~\ref{numeric-stabustab} corresponds to the first and second
lines of initial data above, respectively. The time step for the
simulations was $\Delta t\approx.04453$. The simulations were stopped
if $r$ exceeded $.5$, corresponding to a vehicle configuration which
is skewed $30$~degrees to the vertical.

At the top left, instability is indicated in the gap because large
deviations from the equilibrium are observed for sufficiently long
time. At the bottom left, stability is indicated in the EM-region because of the decreasing deviation from the
equilibrium for decreasing initial conditions, irrespective of the
elapsed simulation time. At the right, energy dissipation is
demonstrated by plotting the energy minus the initial energy
($\Delta\-2 H$) against time.

%
\subsubsection{Transitions}
\label{sec:numericTransitions}
\begin{figure}[t]\setlength{\unitlength}{1in}
\begin{picture}(0,3.25)(-.2,-.025)
\put(.1,1.85){\makebox(0,0)[bl]{\includegraphics{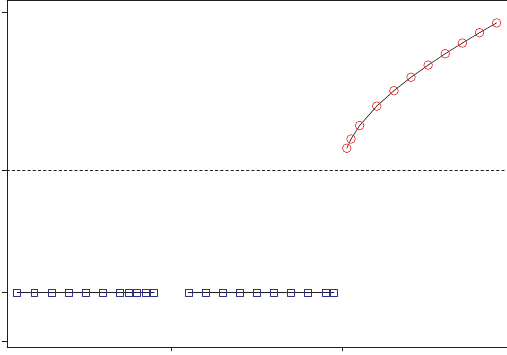}}}
\put(4.9,1.85){\makebox(0,0)[br]{\includegraphics{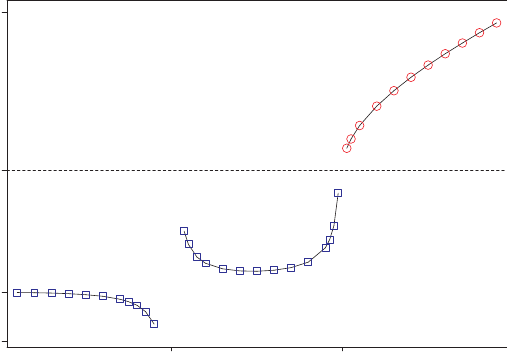}}}
\put(.1,.1){\makebox(0,0)[bl]{\includegraphics{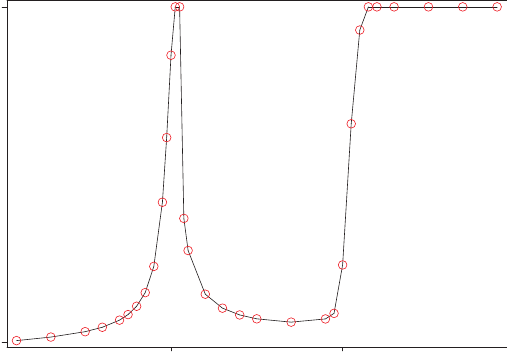}}}
\put(4.9,.1){\makebox(0,0)[br]{\includegraphics{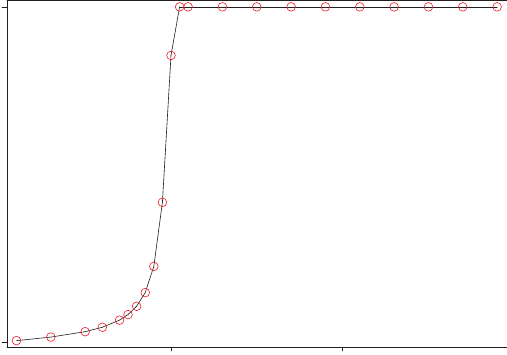}}}
\OrnamentLadybird{4}{3}{.3}{30}
\put(.785,.075){\makebox(0,0)[tc]{\footnotesize$1$}}
\put(1.4725,.075){\makebox(0,0)[tc]{\footnotesize$2$}}
\put(0.025,.8){\makebox(0,0)[cr]{\footnotesize$r$}}
\put(2.075,.025){\makebox(0,0)[tc]{\footnotesize$P_e$}}
\put(.075,.15){\makebox(0,0)[cr]{\footnotesize$0$}}
\put(.075,1.48){\makebox(0,0)[cr]{\footnotesize$.5$}}
\put(2.075,1.775){\makebox(0,0)[tc]{\footnotesize$P_e$}}
\put(.785,1.825){\makebox(0,0)[tc]{\footnotesize$1$}}
\put(1.4725,1.825){\makebox(0,0)[tc]{\footnotesize$2$}}
\put(.075,1.9){\makebox(0,0)[cr]{\footnotesize$-2\times10^{-6}$}}
\put(.075,2.1){\makebox(0,0)[cr]{\footnotesize$0$}}
\put(.075,2.6){\makebox(0,0)[cr]{\footnotesize$5\times10^{-6}$}}
\put(0.075,3.2){\makebox(0,0)[cr]{\footnotesize$1.3$}}
\put(4.86,1.775){\makebox(0,0)[tc]{\footnotesize$P_e$}}
\put(3.56,1.825){\makebox(0,0)[tc]{\footnotesize$1$}}
\put(4.2475,1.825){\makebox(0,0)[tc]{\footnotesize$2$}}
\put(2.85,1.9){\makebox(0,0)[cr]{\footnotesize$-2\times10^{-6}$}}
\put(2.85,2.1){\makebox(0,0)[cr]{\footnotesize$0$}}
\put(2.85,2.6){\makebox(0,0)[cr]{\footnotesize$5\times10^{-6}$}}
\put(2.85,3.2){\makebox(0,0)[cr]{\footnotesize$1.3$}}
\put(3.56,.075){\makebox(0,0)[tc]{\footnotesize$1$}}
\put(4.2475,.075){\makebox(0,0)[tc]{\footnotesize$2$}}
\put(2.80,.8){\makebox(0,0)[cr]{\footnotesize$r$}}
\put(4.86,.025){\makebox(0,0)[tc]{\footnotesize$P_e$}}
\put(2.85,.15){\makebox(0,0)[cr]{\footnotesize$0$}}
\put(2.85,1.48){\makebox(0,0)[cr]{\footnotesize$.5$}}
\end{picture}
\caption{\label{numeric-transition}\it Destruction of KAM stability of
an axisymmetric relative equilibria in the gap between $C_1$ and $C_2$
as a result of adding small dissipation. Left: no dissipation, right:
added dissipation. Top: plots the real parts of the spectrum of nearby
persisting relative equilibria. The scale on the lower parts of the
graphs differs from the scale of the upper parts so that the sign
change at $C_1$ is visible.  The computation was done with the program
MAPLE at 18~digits of accuracy. Bottom: large $r$ indicates
instability.}
\end{figure}
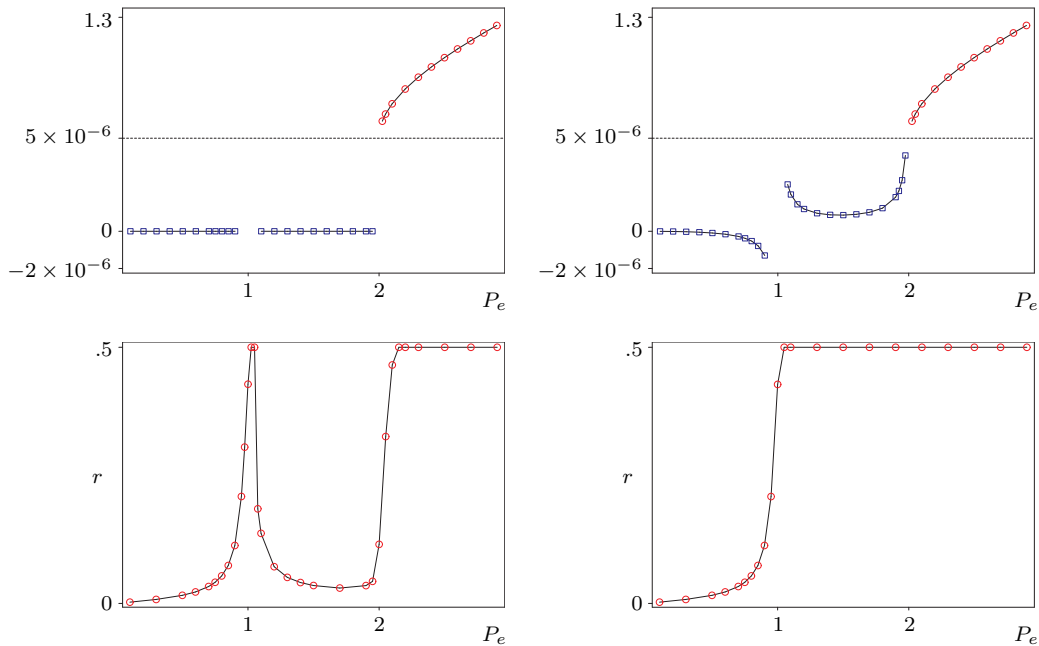
On the left bottom, Figure~\ref{numeric-transition} shows the
maximum value of~$r$ over $1.2\times10^6$ periods, for
\[
\epsilon=0,\quad q_1=0.0125,\quad \nu^a_1=0.0025.
\]
$\Delta t$ was adjusted to $40$ time steps per period. On the right is
the same except for $\epsilon=.05$. For the no-dissipation runs on the
left, stability in the gap between $C_1$ and $C_2$ is indicated by the
small maximum values for $r$. On the right, large deviations from the
equilibrium occur over the entire gap between $C_1$ and $C_2$.

The peak at $C_1$ in the left graph indicates loss of stability at
that transition, even in the purely Hamiltonian context. This is due
to the presence of a zero eigenvalue in the linearization
(see~\eqref{eq:redevals} when $F_q=0$). At $C_1$, KAM~stability as
discussed in Sections~\ref{sec:KAMtheory}
and~\ref{subsection-KAM-stability} is not present because it requires
perturbation from an elliptic equilibrium but the equilibrium is not
elliptic, it has a $0$~eigenvalue.

The equilibria of the $\dot w$-equation~\eqref{eq:simulatedthis} at
$q=p=0$ and vertical momentum $\nu$ persist to nearby nonvertical
momenta.  As further evidence of dissipation induced instability, we
have computed the corresponding spectrum of the linearization of these
equilibria, which were found numerically by Newton's method with start
at $q=p=0$.  As shown in the top of Figure~\ref{numeric-transition},
at zero dissipation, we observe zero real parts in the spectrum
throughout both the EM-region and the gap. For nonzero dissipation,
the spectrum splits and has negative real parts in the EM-region and
positive real parts in the gap. The real parts in the gap are small as
compared to the real parts after the Hopf bifurcation, necessitating
the two vertical scales in the Figure. 

\EndLadybird
\end{document}